\documentclass{article}
\usepackage{amsfonts}
\usepackage{amsmath}
\usepackage{graphicx}
\usepackage{epsfig}

\begin{document}

\title{Extending canonical Monte Carlo methods}
\author{L. Velazquez$^{1,2}$ and S. Curilef$^{2}$ \\
$^{1}$Departamento de F\'{\i}sica, Universidad de Pinar del R\'{\i}o, \\
Mart\'{\i} 270, Esq. 27 de Noviembre, Pinar del R\'{\i}o, Cuba.\\
$^{2}$Departamento de F\'{\i}sica, Universidad Cat\'{o}lica del Norte, \\
Av. Angamos 0610, Antofagasta, Chile.\\
E-mail: lvelazquez@ucn.cl and scurilef@ucn.cl}
\maketitle
\tableofcontents

\begin{abstract}
In this work, we discuss the implications of a recently obtained
equilibrium fluctuation-dissipation relation on the extension of the
available Monte Carlo methods based on the consideration of the
Gibbs canonical ensemble to account for the existence of an
anomalous regime with negative heat capacities $C<0$. The resulting
framework appears as a suitable generalization of the methodology
associated with the so-called \textit{dynamical ensemble}, which is
applied to the extension of two well-known Monte Carlo methods: the
Metropolis importance sample and the Swendsen-Wang clusters
algorithm. These Monte Carlo algorithms are employed to study the
anomalous thermodynamic behavior of the Potts models with many spin
states $q$ defined on a $d$-dimensional hypercubic lattice with
periodic boundary conditions, which successfully reduce the
exponential divergence of decorrelation time $\tau$ with the
increase of the system size $N$ to a weak power-law divergence
$\tau\propto N^{\alpha}$ with $\alpha\approx0.2$ for the particular
case of the 2D 10-state Potts model.
\end{abstract}

\section{Introduction}

In the present work, we shall not propose new Monte Carlo (MC)
methods based on the equilibrium distributions of Statistical
Mechanics. On the contrary, we shall discuss how the available MC
methods based on the consideration of the Gibbs canonical ensemble:
\begin{equation}
dp\left(  E\left\vert \beta_{B}\right.  \right)  =Z\left(
\beta_{B}\right)
^{-1}\exp\left(  -\beta_{B}E\right)  \Omega\left(  E\right)  dE \label{Gibbs}%
\end{equation}
could be extended by using a minimal, but crucial modification in
their schemes to account for the existence of an anomalous regime
with \textit{negative heat capacities} $C<0$ \cite{pad,Lyn3,gro1,gro
na,moretto,Dagostino}. This fact avoids the incidence of the
so-called \textit{super-critical slowing down}, a dynamical anomaly
that significantly affects the efficiency of large-scale canonical
MC simulations \cite{mc3}.

Our proposal follows as a direct application of the recently
obtained fluctuation-dissipation relation \cite{vel-tur,vel-jstat}:
\begin{equation}
C=\beta^{2}\left\langle \delta E^{2}\right\rangle +C\left\langle
\delta \beta_{\omega}\delta E\right\rangle , \label{tur}
\end{equation}
which involves the heat capacity $C$ of a given system in an
equilibrium situation where the inverse temperature $\beta_{\omega}$
of a certain environment exhibits correlated fluctuations with the
system internal energy $E$ as a consequence of their mutual
thermodynamic interaction\footnote{Along this work, Boltzmann
constant is assumed to be $k_{B}\equiv1$.}. Eq.(\ref{tur}) accounts
for the realistic possibility that the internal state of the system
acting as \textit{environment} could be affected by the presence of
the system under study, which is a fact \textit{a priory}
disregarded by the consideration of the Gibbs canonical ensemble
(\ref{Gibbs}), where the inverse temperature $\beta_{\omega}$ of the
environment exhibits a constant value $\beta_{B}$ because its heat
capacity $C_{\omega}$ is practically infinite. Obviously,
Eq.(\ref{tur}) is just a suitable extension of the well-known
relation:
\begin{equation}
C=\beta^{2}\left\langle \delta E^{2}\right\rangle \label{can}%
\end{equation}
between the heat capacity $C$ and the energy fluctuations derived
from the canonical ensemble (\ref{Gibbs}). While the canonical
result (\ref{can}) only admits macrostates with positive heat
capacities $C>0$, it is easy to verify that the fluctuation relation
(\ref{tur}) is compatible with the presence of macrostates having
negative heat capacities $C<0$. This last conclusion is the
fundamental ingredient considered in this work for allowing a direct
extension of some MC algorithms based on the Gibbs canonical
ensemble (\ref{can}) in order to account for a regime with $C<0$. As
discussed elsewhere \cite{pad,Lyn3,gro1,gro na,moretto,Dagostino},
this kind of anomaly in the caloric curve appears to be associated
with the occurrence of a discontinuous (first-order) phase
transition in finite short-range interacting systems, as well as
systems with long-range interactions such as astrophysical systems.

This work is organized into sections as follows: first, we shall
discuss in section \ref{proposal} how the present ideas can be
considered to extend the available canonical MC methods; afterwards,
we shall apply these arguments in section \ref{examples} to extend
two well-known canonical MC algorithms: the \textit{Metropolis
importance sample} \cite{metro,Hastings} and the
\textit{Swendsen-Wang cluster algorithm} \cite{SW,pottsm,wolf}, in
their application to the study of anomalous macrostates present in
the thermodynamic description of the \textit{q}-states Potts models
defined on a $d$-dimensional hypercubic lattice; finally, concluding
remarks are presented in section \ref{conclusions}.

\section{The proposal\label{proposal}}
\subsection{Overview}
For convenience, let us begin the present discussion by reviewing
the most important results related to the fluctuation theorem
(\ref{tur}). Our analysis starts from the consideration of the
following generic energy distribution function \cite{vel-tur}:
\begin{equation}
dp_{\omega}\left(  E\right)  =\omega\left(  E\right)  \Omega\left(
E\right) dE,
\end{equation}
where $\omega\left(  E\right)  $ is a probabilistic weight that
considers the thermodynamic influence of a certain environment. The
above work hypothesis admits the canonical weight:
\begin{equation}\label{can.weight}
\omega_{c}\left( E\right)  =Z\left(  \beta _{B}\right)
^{-1}\exp\left( -\beta_{B}E\right)
\end{equation}
as a relevant but particular case when the environment is just a
thermal bath having an infinite heat capacity. This general
situation could be implemented with the help of a Metropolis Monte
Carlo simulation by using the transition probability:
\begin{equation}
W_{\omega}\left(  E\rightarrow E+\delta E  \right) =\min\left\{
\frac {\omega\left(  E+\delta E\right)  }{\omega\left( E\right)
},1\right\} .\label{T1}
\end{equation}
Since the energy thermal fluctuations $\left\vert \delta
E\right\vert $ are small when the system size is sufficiently large,
Eq.(\ref{T1}) can be rewritten in a \textit{canonical} fashion as
follows:
\begin{equation}\label{T2}
W_{\omega}\left(  E\rightarrow E+\delta E  \right) \simeq\min\left\{
\exp\left[  -\beta_{\omega}\left(  E\right) \delta E\right]
,1\right\}  ,
\end{equation}
where $\beta_{\omega}\left(  E\right)  $ is hereafter referred to as
the \textit{inverse temperature} of the environment:
\begin{equation}
\beta_{\omega}\left(  E\right)  =\frac{1}{T_{\omega}\left(  E\right)  }%
=-\frac{\partial}{\partial E}\log\omega\left(  E\right)  .\label{eff.temp}%
\end{equation}

The density of states $\Omega\left(  E\right)  $ is related to the
system entropy $S\left(  E\right)  \,$\ as $\Omega\left(  E\right)
\equiv C\exp\left[  S\left(  E\right)  \right]  $, which allows us
to obtain the system temperature $T$ by using the thermodynamic
relation:
\begin{equation}
\beta\left(  E\right)  =\frac{1}{T\left(  E\right)  }=\frac{\partial
S\left(
E\right)  }{\partial E}.\label{sys.temp}%
\end{equation}
Eqs.(\ref{eff.temp}) and (\ref{sys.temp}) can be combined to express
the \textit{inverse temperature difference} $\eta$ as follows:
\begin{equation}
\eta\left(  E\right)  =\beta_{\omega}\left(  E\right)  -\beta\left(
E\right)
\equiv-\frac{\partial}{\partial E}\log\rho\left(  E\right)  ,\label{temp.diff}%
\end{equation}
with $\rho\left(  E\right)  =\omega\left(  E\right)  \Omega\left(
E\right)  $ being the density of probability. This last
representation (\ref{temp.diff}) is very useful to obtain two
remarkable thermodynamic relations. The first one involves the
statistical expectation value $\left\langle \eta\right\rangle $, and
its calculation reads as follows:
\begin{align}
\left\langle \eta\right\rangle  &
=\int_{E_{\inf}}^{E_{\sup}}\eta\left( E\right)  \rho\left(  E\right)
dE=-\int_{E_{\inf}}^{E_{\sup}}\frac{\partial
}{\partial E}\rho\left(  E\right)  dE\nonumber\\
&  =-\left.  \rho\left(  E\right)  \right\vert _{E_{\inf}}^{E_{\sup}%
}=0,\label{exp.value.1}%
\end{align}
while the second one considers the correlation function
$\left\langle E\eta\right\rangle $:
\begin{align}
\left\langle E\eta\right\rangle  &
=\int_{E_{\inf}}^{E_{\sup}}E\eta\left( E\right)  \rho\left( E\right)
dE=-\int_{E_{\inf}}^{E_{\sup}}E\frac{\partial
}{\partial E}\rho\left(  E\right)  dE\nonumber\\
&  =\int_{E_{\inf}}^{E_{\sup}}\rho\left(  E\right)dE  =1.\label{exp.value.2}%
\end{align}
Here, we have taken into account the vanishing of the density of
probability $\rho\left(  E\right)  $ and its first derivative
$\partial\rho\left( E\right)  /\partial E$ at the maximum $E_{\sup}$
and minimum $E_{\inf}$ values of the system energy, as well as the
normalization condition.

The vanishing of the expectation value $\left\langle
\eta\right\rangle $, Eq.(\ref{exp.value.1}), is simply the known
\textit{thermal equilibrium condition}:
\begin{equation}\label{thermal.eq}
\left\langle \frac{1}{T_{\omega}}\right\rangle \equiv\left\langle \frac{1}%
{T}\right\rangle ,
\end{equation}
the mathematical form of which clearly indicates that the
equalization of temperature expressed by the \textit{Zeroth
Principle of Thermodynamics} takes place in an average sense.
Eq.(\ref{exp.value.2}) can be rewritten as a rigorous fluctuation
relation by using the identity $\left\langle \delta
E\delta\eta\right\rangle \equiv\left\langle E\eta\right\rangle
-\left\langle E\right\rangle \left\langle \eta\right\rangle $:
\begin{equation}\label{fluct.gen}
\left\langle \delta E\delta\left(
\frac{1}{T_{\omega}}-\frac{1}{T}\right) \right\rangle =1.
\end{equation}
By using the \textit{Schwartz inequality} $\left\langle
AB\right\rangle ^{2}\leq\left\langle A^{2}\right\rangle \left\langle
B^{2}\right\rangle $, this last result can be rephrased as:
\begin{equation}\label{unc}
\Delta E\Delta\left(  \frac{1}{T_{\omega}}-\frac{1}{T}\right) \geq1,
\end{equation}
where $\Delta x\equiv\sqrt{\left\langle \delta x^{2}\right\rangle}$.
Finally, by substituting the first-order approximation:
\begin{equation}\label{GApp}
\delta\left(\frac{1}{T}\right)\simeq-\frac{1}{T^{2}C}\delta E,
\end{equation}
into Eq.(\ref{fluct.gen}), with $C=dE/dT$ being the heat capacity,
one obtains the fluctuation-dissipation relation (\ref{tur}).

The fluctuation-dissipation relation (\ref{tur}) can be rewritten as
follows:
\begin{equation}
C\left(1-\left\langle \delta \beta_{\omega}\delta
E\right\rangle\right)=\beta^{2}\left\langle \delta
E^{2}\right\rangle .
\end{equation}
Since the right-hand side of this expression is always nonnegative,
it is easy to see that the presence of macrostates with positive
heat capacities $C>0$ demands that the correlation function
$\left\langle \delta \beta_{\omega}\delta E\right\rangle$ obey the
constraint:
\begin{equation}\label{fluc.constrain1}
\left\langle \delta \beta_{\omega}\delta E\right\rangle<1.
\end{equation}
Clearly, such a condition is fulfilled by the Gibbs canonical
ensemble (\ref{Gibbs}), where $\delta \beta_{\omega}\equiv 0$.
However, the existence of macrostates having negative heat
capacities $C<0$ can be only observed provided that the constraint:
\begin{equation}\label{fluc.constrain2}
\left\langle \delta \beta_{\omega}\delta E\right\rangle>1
\end{equation}
holds. Thus, any attempt to impose the canonical condition $\delta
\beta_{\omega}\rightarrow 0$ is always accompanied with a
progressive increase of the energy fluctuations $\delta
E\rightarrow\infty$, which leads to the thermodynamic instability
and inaccessibility of such anomalous macrostates.

The simplest way to guarantee the existence of non-vanishing
correlated fluctuations $\left\langle \delta \beta_{\omega}\delta
E\right\rangle\neq0$ is achieved by considering an environment with
a \textit{finite heat capacity} $C_{\omega}$. Here, the inverse
temperature fluctuations $\delta \beta_{\omega}$ can be expressed in
terms of the amount of energy $\delta E$ released or absorbed by the
system in turn its equilibrium value:
\begin{equation}\label{GappE}
\delta\beta_{\omega}\simeq\beta^{2}\frac{1}{C_{\omega}}\delta E,
\end{equation}
where we have considered the thermal equilibrium condition
$\beta_{\omega}=\beta$. By substituting this last expression into
fluctuation-dissipation relation (\ref{tur}), one obtains:
\begin{equation}
\frac{CC_{\omega}}{C+C_{\omega}}=\beta^{2}\left\langle \delta
E^{2}\right\rangle .
\end{equation}
Since the right-hand side of this last expression is always
nonnegative, the thermodynamic stability of macrostates with
negative heat capacity $C<0$ demands the applicability of the
following constraint \cite{vel-jstat}:
\begin{equation}\label{thir.const}
0<C_{\omega}<|C|.
\end{equation}
Remarkably, this result was also obtained in the past by Thirring
\cite{Thirring}.

\begin{figure}
[t]
\begin{center}
\includegraphics[
height=2.6in, width=3.4in
]%
{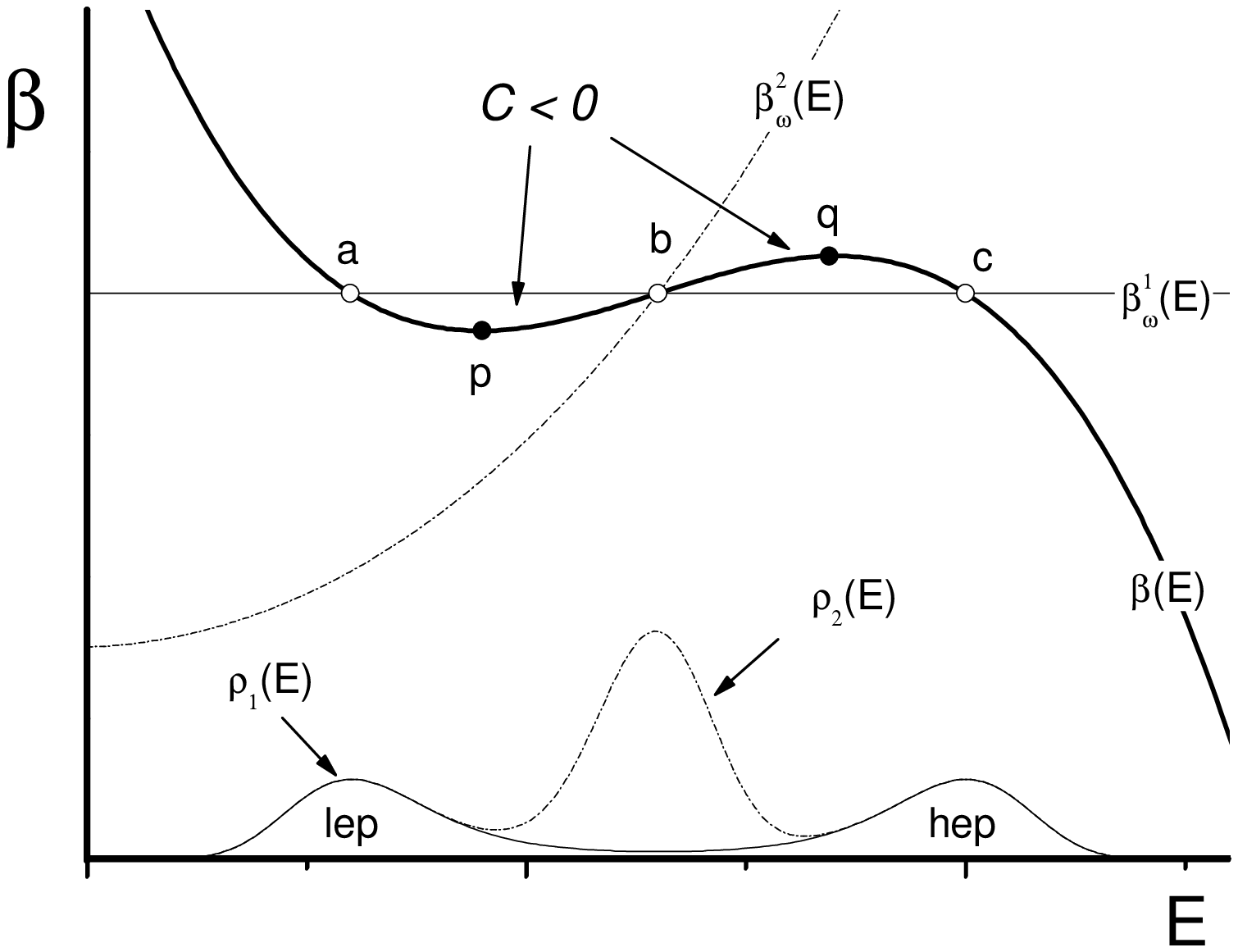}%
\caption{Schematic behavior of the microcanonical caloric
$\beta\left( E\right)  =\partial S\left( E\right) /\partial E$ of a
finite short-range interacting system undergoing a first-order phase
transition between a low (lep) and a high (hep) energy phase. The
density of probabilities $\rho_{1}\left( E\right)$ and
$\rho_{2}\left( E\right)$ are associated with the thermal contact of
this system with certain environments characterized by the inverse
temperatures $\beta^{1}_{\omega}\left( E\right)$ and
$\beta^{2}_{\omega}\left( E\right)$ respectively.}
\label{caricature.eps}%
\end{center}
\end{figure}

The above consequences are illustrated in detail in
FIG.\ref{caricature.eps}. We show here the typical backbending
behavior of the microcanonical caloric curve $\beta(E)$ of a finite
short-range interacting system undergoing a first-order phase
transition \cite{gro1}, where the points within the energetic region
$E_{p}<E<E_{q}$ represent anomalous macrostates with negative heat
capacities. The density of probability $\rho\left(E\right)$
corresponds to a situation where this system is put in thermal
contact with a certain environment characterized by the inverse
temperature $\beta_{\omega}(E)$. The maxima and minima of the
distribution function $\rho\left(E\right)$ are determined from the
thermal equilibrium condition $\beta\left( E\right)
=\beta_{\omega}\left( E\right)$, that is, the intersection points
between these inverse temperature dependencies. For convenience, we
have shown here two relevant cases.

The first case corresponds to the equilibrium situation associated
with the Gibbs canonical ensemble (\ref{Gibbs}), where the inverse
temperature dependence $\beta_{B}^{1}\left( E\right)$ remains at the
constant value $\beta_{B}$ despite the underlying energy
interchange. It should be noticed that the thermal equilibrium
condition is fulfilled by three points $\left(
E_{a},E_{b},E_{c}\right) $ when the inverse temperature $\beta_{B}$
is within the interval $\left(\beta _{p},\beta_{q}\right)$. If this
is the case, the energy distribution function
$\rho_{1}\left(E\right) $ is \textit{bimodal}, where the points
$E_{a}$ and $E_{c}$ determine the positions of its peaks (local
maxima), while the intermediate point $E_{b}$ determines the
position of its local minimum. One can verify that the local minimum
$E_{b}$ always belongs to the anomalous region with $C<0$, while the
maxima $\left( E _{a},E_{c}\right)  $ are located within the regions
with $C>0$. The accessibility of such points behaves with the
increase of the system size $N$ as $\rho_{1}\left( E_{a,c}\right)
\propto e^{\alpha _{a,c}N}$ and $\rho_{1}\left( E_{b}\right) \propto
e^{-\alpha _{b}N}$, where $\alpha_{a,b,c}>0$. Consequently,
anomalous macrostates with $C<0$ becomes \textit{practically
inaccessible} within the canonical ensemble (\ref{Gibbs})\ when $N$
is sufficiently large. The existence of such a \textit{hidden
region} is the origin of the latent heat $q_{L}$ necessary for the
conversion of one phase into the other during the phase transition,
as well as for the \textit{ensemble inequivalence} between the
microcanonical and canonical description \cite{gro1}.

A multi-modal character of the density of probability
$\rho_{1}\left( E\right)  $ in the framework of MC simulations based
on the canonical ensemble (\ref{Gibbs}) leads to the occurrence of
the \textit{super-critical slowing down} \cite{mc3}: an exponential
divergence of correlation times with the increasing of the system
size, $\tau\propto \exp\left(  \lambda N\right)  $. In other words,
this phenomenon manifests as an \textit{effective trapping} of the
system energy within any of the coexisting peaks of the distribution
function $\rho_{1}\left(  E \right)  $, due to the probability $T$
for the occurrence of a large energy fluctuation that allows a
transition towards a neighboring peak decreases exponentially with
the increase of the system size $N$, $T\simeq\rho_{1}\left(
E_{b}\right) /\rho_{1}\left( E_{a}\right) \propto e^{-\lambda N}$.
Thus, the characteristic timescale for such a transition is given by
$\tau\propto 1/T\sim\exp\left(  \lambda N\right)  $. Because the
canonical averages should consider the contribution of all these
coexisting peaks, the relaxation timescales of such expectation
values also exhibit an exponential growth with the increasing of
$N$.

The second case shown in FIG.\ref{caricature.eps} corresponds to a
situation where the system is put in thermal contact with an
environment having a finite heat capacity. By choosing appropriately
the environment and its internal conditions, in particular, the
applicability of Thirring's constraint (\ref{thir.const}), the
corresponding inverse temperature $\beta_{\omega}^{2}\left( E\right)
$ can ensure the existence of only one intersection point with the
microcanonical caloric curve $\beta\left(  E\right)  $ of the system
under study. Even, such a point could be located within the
anomalous region with $C<0$, e.g., the unstable macrostate $E_{b}$.
Since the energy distribution function $\rho_{2}\left( E\right)  $
is monomodal, the phenomenon of super-critical slowing down cannot
be present when such a physical situation is simulated by using a
suitable MC method.

\subsection{Application in Monte Carlo methods}
In the multicanonical MC method and its variants \cite{mc3}, the
main aim is to obtain of the density of states $\Omega(E)$, or,
equivalently, the microcanonical entropy $S(E)$. Such a goal can be
achieved through a direct MC calculation of the energy distribution
function $\rho(E)=\omega(E)\Omega(E)$, which can be inverted to
express the system entropy and the canonical expectation values
$\left\langle A\right\rangle_{\beta}$ as follows:
\begin{equation}
 S(E)=\log\left[\rho(E)\right]-\log\left[\omega(E)\right]+cte,
\end{equation}
\begin{equation}
  \left\langle A\right\rangle_{\beta}=\left[\int A(E)
\frac{\rho(E)}{\omega(E)}e^{-\beta_{B} E}dE\right]\left[\int
\frac{\rho(E)}{\omega(E)}e^{-\beta_{B} E}dE\right]^{-1}.
\end{equation}
The mathematical form of the probabilistic weight, $\omega(E)$, can
be conveniently proposed before performing the MC simulation
\cite{Viana}, although the most common strategy is to carry out its
iterative reconstruction through several preliminary MC runs to
obtain a \textit{flat} histogram $\rho(E)\sim const$ within the
energy interval of interest \cite{BergM,WangLandau}. This latter
alternative allows to enhance rare events, which is particularly
useful to avoid the super-critical slowing down near to first-order
phase transitions \cite{mc3}. This kind of MC simulation allows the
acquisition of the thermodynamic information in a wide energy
interval by performing a single MC run. A clear disadvantage is that
a substantial fraction of computational resources should be consumed
to find an optimal probabilistic weight $\omega(E)$. Moreover, the
acquisition of the microcanonical caloric curve $\beta(E)=\partial
S(E)/\partial E$ as well as of the heat capacity
$C(E)=-\beta^{2}(E)\left[\partial^{2}S(E)/\partial
E^{2}\right]^{-1}$ from a numerical differentiation of the entropy
$S(E)$ is a procedure that enhances the unavoidable statistical
errors involved in the MC calculation of the energy distribution
$\rho(E)$ and/or the probabilistic weight $\omega(E)$.

Clearly, it is desirable to implement some type of methodology that
allows a direct estimation of the caloric curve $\beta(E)$ and of
the heat capacity $C(E)$ without considering a numerical
differentiation of the system entropy $S(E)$. A simple strategy is
provided by considering the thermal contact of the system with an
environment characterized by a finite heat capacity, as in the
second case illustrated in FIG.\ref{caricature.eps}. It should be
noticed here that the system energy $E$ and the inverse temperature
$\beta_{\omega}$ of the environment undergo small thermal
fluctuations around their expectation values $\left\langle E
\right\rangle $ and $\left\langle \beta_{\omega}\right\rangle $,
which provide a suitable estimation of the intersection point
$(E_{e},\beta_{e})$ derived from the condition of thermal
equilibrium $\beta_{\omega}(E_{e})=\beta(E_{e})\equiv\beta_{e}$:
\begin{equation}\label{ave.inter}
E_{e}=\left\langle E \right\rangle,~\beta_{e}=\left\langle
\beta_{\omega} \right\rangle.
\end{equation}

In essence, Eq.(\ref{ave.inter}) expresses the procedure associated
with the MC method of the so-called \textit{dynamical ensemble}
proposed by Gerling and H\"{u}ller \cite{Gerling}, which is defined
by a probabilistic weight $\omega(E)$ with a power-law shape:
\begin{equation}\label{GHens}
\omega_{DE}(E)=N_{0}\left(E_{T}-E\right)^{B},
\end{equation}
whose corresponding inverse temperature is given by:
\begin{equation}
\beta_{\omega}(E)=B/(E_{T}-E).
\end{equation}
The fluctuation relation (\ref{tur}) enables the introduction of
several improvements for this kind of MC calculations to estimate
the thermo-statistical properties of a system within the
microcanonical ensemble by using modified canonical MC algorithms.
In fact, one can also obtain the heat capacity $C_{e}=C(E_{e})$ at
the interception point from the fluctuating behavior of the system
as follows:
\begin{equation}\label{heat.fluc}
C_{e}=\frac{\beta_{e}^{2}\left\langle\delta E^{2}
\right\rangle}{1-\left\langle \delta \beta_{\omega}\delta
E\right\rangle}.
\end{equation}
Thus, one is able to obtain a suitable estimation of any point on
the microcanonical caloric curve $\beta\left(E\right)$ as well as
the heat capacity $C(E)$ regardless of whether its character
positive or negative.

\subsubsection{The linear ansatz and its optimization}
Generally speaking, the specific mathematical form of the
probabilistic weight $\omega(E)$ or its corresponding inverse
temperature $\beta_{\omega}(E)$ is unimportant as long as the
following conditions applied:
\begin{enumerate}
    \item The inverse temperature $\beta_{\omega}(E)$ of the environment
must ensure the existence of only one intersection point with the
caloric curve $\beta(E)$, that is, the existence of only one peak in
the energy distribution function $\rho(E)$.
    \item The size $N$ of the system under analysis should be
sufficiently large to guarantee the validity of the first-order
approximation (\ref{GApp}) considered for deriving the
fluctuation-dissipation relation (\ref{tur}) from the rigorous
fluctuation relation (\ref{fluct.gen}).
\end{enumerate}
As already depicted in FIG.\ref{caricature.eps}, the energy
distribution function $\rho(E)$ has the shape of a bell curve,
usually approximated by a Gaussian distribution. Since the system
exhibits small thermal fluctuations $\Delta E/|E|\sim1/\sqrt{N}$,
only local properties of the inverse temperature dependence
$\beta_{\omega}(E)$ close to the equilibrium point $E_{e}$ are
significant; hence, one can obtain the same practical results by
using different mathematical dependencies for the inverse
temperature $\beta_{\omega}(E)$.

The simplest mathematical form is provided by restricting to the
first-order power expansion of the inverse temperature dependence
$\beta_{\omega}(E)$ around the intersection point:
\begin{equation}\label{ansatz.linear}
\beta_{\omega}(E)=\beta_{e}+\lambda\delta E/N,
\end{equation}
which considers a linear coupling of the environment inverse
temperature $\beta_{\omega}$ with the thermal fluctuations of the
system energy $\delta E=E-E_{e}$. Remarkably, such a linear ansatz
(\ref{ansatz.linear}) is equivalent to the so-called
\textit{Gaussian ensemble} \cite{ChallaH}:
\begin{equation}\label{GEns}
\omega_{GE}(E)=A\exp\left[-\beta_{e}E-\frac{1}{N}\lambda(E-E_{e})^{2}\right]
\end{equation}
proposed by Hetherington \cite{Hetherington}, which approaches in
the limit $\lambda\rightarrow+\infty$ to the microcanonical ensemble
$\omega_{ME}(E)=A\delta(E-E_{e})$.

Hypothesis (\ref{ansatz.linear}) ensures the existence of
non-vanishing correlated fluctuations $\left\langle \delta
\beta_{\omega}\delta E\right\rangle\neq0$ when the coupling constant
$\lambda\neq0$. By using the fluctuation-dissipation relation
(\ref{tur}) and the ansatz (\ref{ansatz.linear}), one can obtain the
following expression for the heat capacity:
\begin{equation}\label{cap.fluct}
C_{e}=\frac{\beta_{e}^{2}(\Delta E)^{2}}{1-\lambda(\Delta E)^{2}/N},
\end{equation}
which can be rewritten to obtain the energy and inverse temperature
dispersions, $\Delta E$ and $\Delta \beta_{\omega}$, as follows:
\begin{equation}\label{dispersions}
(\Delta E)^{2}=\frac{N}{\beta_{e}^{2}N/C_{e}+\lambda},~(\Delta
\beta_{\omega})^{2}=\frac{1}{N}\frac{\lambda^{2}}{\beta_{e}^{2}N/C_{e}+\lambda},
\end{equation}
where $\Delta x\equiv\sqrt{\left\langle\delta x^{2}\right\rangle}$.
The inverse temperature dispersion $\Delta\beta_{\omega}$ decreases
with the increase of the system size $N$ as
$\Delta\beta_{\omega}=|\lambda|\Delta E/N\sim1/\sqrt{N}$. In the
limit $N\rightarrow+\infty$, the thermal fluctuations disappear and
the present equilibrium situation becomes equivalent to the
microcanonical ensemble provided that the following condition:
\begin{equation}\label{CEq}
\beta_{e}^{2}N/C_{e}+\lambda>0
\end{equation}
holds. According to the first-order approximation (\ref{GappE}), the
coupling constant $\lambda$ can be related to the heat capacity
$C_{\omega}$ of the environment as
$\lambda=\beta_{e}^{2}N/C_{\omega}$, hence, the stability condition
(\ref{CEq}) is fully equivalent to Thirring's constraint
(\ref{thir.const}).

At first glance, it is desirable to maximally reduce the thermal
dispersions of the system energy $E$ and its inverse temperature
$\beta=1/T$ indirectly derived from the environment inverse
temperature $\beta_{\omega}=1/T_{\omega}$. However, the inequality
of Eq.(\ref{unc}) imposes an important limitation on the precision
of such a kind of measuring process: a reduction of the thermal
uncertainties $\Delta (1/T_{\omega}-1/T)$ affecting the temperature
equalization provokes an increasing of the system energy
fluctuations $\Delta E$, and vice versa. As previously discussed
\cite{vel-tur,vel-jstat}, Eq.(\ref{unc}) accounts for the existence
of some type of \textit{complementary character} between
thermodynamic quantities of energy and temperature. In consequence,
one must assume the existence of non-vanishing thermal uncertainties
for the system energy $E$ and its inverse temperature $\beta$.

According to expressions (\ref{dispersions}), the increase of the
coupling constant $\lambda$ allows us to reduce the energy
dispersion $\Delta E$, but its value should not be excessively large
because its increasing also leads to the increasing of the inverse
temperature dispersion $\Delta\beta_{\omega}$. A simple criterion to
provide the best value for $\lambda$ is obtained after minimizing
the total dispersion $\Delta_{T} ^{2}=(\Delta E)^{2}/N+N(\Delta
\beta_{\omega})^{2}$. By introducing the \textit{microcanonical
curvature} $\kappa$:
\begin{equation}
\kappa=\kappa(E)=\beta^{2}N/C\equiv-N\partial^{2}S\left(  E\right)
/\partial E^{2}, \label{curvature}%
\end{equation}
which is defined in terms of the second derivative of the entropy
with the opposite sign, the optimal value (opt) for $\lambda$ is
given by:
\begin{equation}
\lambda_{\text{opt}}\left(  \kappa\right)
=\sqrt{\kappa^{2}+1}-\kappa.
\label{optimal}%
\end{equation}
The microcanonical caloric curve $\beta\left(  E\right)$ and the
heat capacity $C\left( E\right)$ derived from Eqs.(\ref{ave.inter})
and (\ref{cap.fluct}) should not depend on the coupling constant
$\lambda$ as long as this latter parameter fulfils condition
(\ref{CEq}) and the system size $N$ is sufficiently large.
Nevertheless, the use of its optimal value (\ref{optimal}) minimizes
the underlying thermal fluctuations, which should also reduce the
number of steps needed to ensure the convergence of a MC run.

\subsubsection{Iterative schemes}

Given a certain dependence of the environment inverse temperature
$\beta^{(i)}_{\omega}(E)$, one can obtain from a MC run suitable
estimations of the system inverse temperature $\beta_{i}$ and its
heat capacity $C_{i}$ at the $i$-th interception point $E_{i}$.
These estimates values can be employed to provide the next
dependence $\beta^{(i+1)}_{\omega}(E)$ to perform an analogous
estimation at the $i+1$-st neighboring point $E_{i+1}$. To fix some
ideas, let us denote by $\varepsilon=E/N$ the system energy per
particle. The $i$-th dependence of the environment inverse
temperature (\ref{ansatz.linear}) considered for the MC calculation
of the $i$-th point $\varepsilon_{i}$ is given by:
\begin{equation}
\beta^{(i)}_{\omega}\left(  \varepsilon\right)
=\bar{\beta}_{i}+\lambda_{i}\left(
\varepsilon-\bar{\varepsilon}_{i}\right)  , \label{variable}%
\end{equation}
where $\bar{\varepsilon}_{i}$ and $\bar{\beta}_{i}$ are some rough
estimates of the expectation values
$\varepsilon_{i}=\left\langle\varepsilon\right\rangle$ and
$\beta_{i}=\left\langle\beta^{(i)}_{\omega}\right\rangle$. The
parameters $(\bar{\varepsilon}_{i},\bar{\beta}_{i},\lambda_{i})$
could be provided in an interactive way by using the curvature
$\kappa_{i}=\beta_{i}^{2}N/C_{i}$ derived from the previous MC
simulation:
\begin{equation}
\bar{\varepsilon}_{i+1}=\bar{\varepsilon}_{i}+\Delta
e,~\bar{\beta}_{i+1} =\bar{\beta}_{i}-\kappa_{i}\Delta
e,~\lambda_{i+1}=\lambda_{\text{opt}}\left(  \kappa_{i}\right),
\label{itter1}
\end{equation}
where $\Delta e$ is a small energy step. The initial value
$\bar{\varepsilon} _{0}$ is assumed to be the expectation value
$\left\langle \varepsilon\right\rangle$ of the energy per particle
obtained from any canonical MC simulation with inverse temperature
$\beta_{B}=\bar{\beta}_{0}$ sufficiently away from the anomalous
region with $C<0$.

The stability of the iterative procedure previously described
depends on the precision of the estimated value of microcanonical
curvature $\kappa$. Of course, alternative iteration schemes are
also possible, e.g., the following scheme:
\begin{equation}
\bar{\varepsilon}_{i+1}=\left\langle \varepsilon\right\rangle_{i}
+\Delta e,~\bar{\beta}_{i+1}=\left\langle
\beta_{\omega}\right\rangle_{i} , \label{itter2}
\end{equation}
which forces with $\Delta e>0$ ($\Delta e<0$) a forward (backward)
motion of the expectation values $\left\langle
\varepsilon\right\rangle $ and $\left\langle
\beta_{\omega}\right\rangle $ along the system caloric curve
$\beta\left( \varepsilon\right) $ whenever the value of coupling
parameter $\lambda$ does not change in a significant way. In
general, the inverse temperature $\beta\left( \varepsilon\right)$ of
a large enough short-range interacting system describes \textit{a
plateau} within the anomalous region with $C<0$, which means that
the corresponding curvature $\kappa$ does not significantly differ
from zero, $\kappa\simeq0$. In such cases, the value
$\lambda\simeq1$ constitutes a suitable approximation within the
energy range containing the anomalous region with $C<0$, while the
value $\lambda=0$ corresponding to the canonical ensemble is a good
choice elsewhere.

\subsubsection{Implementation}

Since the inverse temperature $\beta_{B}$ of the Gibbs canonical
ensemble (\ref{Gibbs}) appears as a driving parameter in the
transition probability $W(X\rightarrow X';\beta_{B})$ of canonical
MC methods, the most general idea to extend this kind of algorithms
is to replace the bath inverse temperature $\beta_{B}$ by a variable
inverse temperature $\beta_{\omega}(E)$,
$\beta_{B}\rightarrow\beta_{\omega}(E)$. Moreover, it is desirable
that the transition probability resulting from such a modification
fulfills the so-called \textit{detailed balance condition}
\cite{mc3}:
\begin{equation}\label{DBal}
p(X)W(X\rightarrow X')=p(X')W(X'\rightarrow X),
\end{equation}
where $p(X)$ is the system distribution function, which is given by
the function $p_{\omega}(X)=\omega\left[E(X)\right]$.

Let us denote the energy varying during the configuration change
$X\rightarrow X'$ as $\delta E=E\left(  X'\right) -E\left( X\right)
$ and its mean value as $E^{m}=\left[  E\left( X\right) +E\left(
X'\right)  \right]  /2$. According to the mean value theorem, one
can express the variation of the function $\log p_{\omega}\left(
X\right)$ by using the environment inverse temperature
$\beta_{\omega}\left(  E\right) $ evaluated at a certain
intermediate energy $E_{\theta}=E^{m}+\theta\delta E$ as follows:
\begin{equation}
\log p_{\omega}\left(  X'\right)-\log p_{\omega}\left( X\right)
=-\beta_{\omega}^{\star}\delta E,\label{meanvalue}
\end{equation}
where $\theta$ is a real parameter with $\left\vert
\theta\right\vert \leq1$\ and
$\beta_{\omega}^{\star}\equiv\beta_{\omega}\left(
E_{\theta}\right)$. By considering the transition probability of any
canonical MC algorithm that obeys the detailed balance condition:
\begin{equation}
\frac{W\left(  X\rightarrow X';\beta_{B}\right)  }{W\left(
X'\rightarrow X;\beta_{B}\right)  }\equiv\exp\left( -\beta_{B}\delta
E\right) ,
\end{equation}
as well as Eq.(\ref{meanvalue}), one can finally obtain:
\begin{equation}
p_{\omega}\left(  X\right)  W\left(  X\rightarrow X';\beta_{\omega
}^{\star}\right)  =p_{\omega}\left( X'\right)  W\left( X'\rightarrow
X;\beta_{\omega}^{\star}\right) .\label{detailed.bal.omega}
\end{equation}
This last result clarifies that given the initial and final system
configurations, $X$ and $X'$, one can always find a certain value
$\beta_{\omega}^{\star}\equiv\beta_{\omega}\left( X,X'\right)  $ of
the bath inverse temperature that fulfils the detailed balance
condition (\ref{DBal}) for the present distribution function
$p_{\omega }\left(  X\right) $. In particular, the exact value of
$\beta^{\star}_{\omega}$ for the Gaussian ensemble (\ref{GEns}) is
the one corresponding to $\theta=0$,
$\beta^{\star}_{\omega}=\beta_{\omega}(E^{m})$.

The main obstacle to perform a direct application of
Eq.(\ref{detailed.bal.omega}) to extend canonical MC methods relies
on the fact that the final system configuration $X^{\prime}$ is
\textit{a priori} unknown in many non-local MC algorithms
\cite{SW,pottsm,wolf,Edwards,Niedermayer,Evertz,Hasenbusch,Dress,Liu}.
Consequently, one should consider some suitable approximation
$\beta_{\omega}^{\ast} =\beta_{\omega}\left(
E_{\theta^{\ast}}\right)  $ of the exact value
$\beta^{\star}_{\omega}$, e.g., the one corresponding to the
environment inverse temperature at the initial system configuration
$X$, $\beta_{\omega}^{i}=\beta_{\omega}\left[E(X)\right]$. To
estimate the error involved in this last approximation, it is
important to take into account that the environment heat capacity
$C_{\omega}$ and the energy change $\delta E$ behave with the
increase of the system size $N$ as $C_{\omega}\sim N$ and $\delta
E\sim N^{\alpha}$. Here, the exponent $\alpha$ ranges from zero for
a local algorithm such as the Metropolis importance sample, up to
$1/2$ for a hypothetical non-local algorithm able to obtain an
effective independent configuration after each MC step\footnote{The
exponent $\alpha=1/2$ follows from the size behavior of the energy
dispersion $\Delta E\sim\sqrt{N}$.}. Consequently, the difference
$\delta_{\beta}=\left\vert
\beta_{\omega}^{i}-\beta_{\omega}^{\star}\right\vert $ merely
constitutes a small size effect:
\begin{equation}\label{small.error}
\delta_{\beta}\sim\left(  \beta_{\omega}^{m}\right)
^{2}\frac{1}{C_{\omega }^{m}}\left\vert \delta E\right\vert \sim
\frac{1}{N^{1-\alpha}}  ,
\end{equation}
where the index \textit{m} indicates that the corresponding
quantities $\beta_{\omega}$ and $C_{\omega}$ have been evaluated at
the energy value $E^{m}$. Since the estimated inverse temperature
$\beta^{i}_{\omega}$ depends on the system energy $E(X)$, the
corresponding distribution function associated with this
approximation $p^{*}_{\omega}(X)$ can also be expressed by a certain
function of the system energy,
$p^{*}_{\omega}(X)=\omega^{*}\left[E(X)\right]$. According to
Eq.(\ref{small.error}), the corresponding inverse temperature:
\begin{equation}
\beta_{\omega^{*}}(E)=-\frac{\partial\log\omega^{*}(E)}{\partial E}
\end{equation}
cannot differ in a significant way from the exact dependence
$\beta_{\omega}(E)$ as long as the system size $N$ be sufficiently
large.

The extended canonical MC methods based on an estimation of the
inverse temperature $\beta^{\star}_{\omega}$ do not fulfill the
detailed balance condition (\ref{DBal}). However, this fact does not
represent any fundamental difficulty since one practically obtains
the same numerical results for the caloric curve $\beta(E)$ and the
heat capacity $C(E)$ with the help of equations (\ref{ave.inter})
and (\ref{cap.fluct}) by using slightly different probabilistic
weight $\omega^{*}(E)$. The only requirement is that the energy
distribution function $\rho(E)$ exhibits a sharp Gaussian profile,
which is simply achieved when the size $N$ of the system under study
is large enough. While the super-critical slowing down of canonical
MC methods observed in systems undergoing a first-order phase
transition becomes more severe as $N$ increases, the errors
associated with all approximations assumed here turn more and more
negligible. This is the reason why the present methodology is
particularly useful for avoiding this type of slow sampling problems
in large-scale MC simulations.

As naturally expected, finite size effects can be significant when
one is also interested to describe systems whose size $N$ is not so
large. If this is the case, it is not only necessary to implement
extended MC schemes that fulfills the detailed balance condition
(\ref{DBal}), but also the inclusion of some finite size corrections
into equations (\ref{ave.inter}) and (\ref{cap.fluct}) employed here
to obtain the microcanonical dependencies $\beta(E)$ and $C(E)$.
Although the complete analysis of these questions is outsize of the
scope of the present paper, we would like to clarify that the most
general way to fulfill the detailed balance condition (\ref{DBal})
after the consideration of an estimated value for the transition
inverse temperature $\beta_{\omega}^{\star}$ is to introduce
\textit{a posteriori} an acceptance probability $w$:
\begin{equation}
w=\min \left\{ 1,\frac{W_{f\rightarrow i}}{W_{i\rightarrow f}}\exp
\left( -\beta_{\omega}^{\star}\delta E\right) \right\} \label{wif}
\end{equation}%
to accept or reject the final configuration $X'$. Here, $%
W_{i\rightarrow f}=W\left[ X\rightarrow X';\beta _{\omega
}^{i}\right] $ and $W_{f\rightarrow i}=W\left[ X'\rightarrow X;\beta
_{\omega }^{f} \right] $, with $\beta _{\omega }^{i}=\beta _{\omega
}(X)$ and $\beta _{\omega }^{f}=\beta _{\omega }(X')$, represent the
transition probabilities of the direct and the reverse process,
respectively. The mathematical forms of which depend on the
particularities of each non-local MC method. 

\section{Application examples\label{examples}}

\subsection{The model}

For the sake of simplicity, let us consider in the present study the
$q$-state
Potts model \cite{pottsm}:%
\begin{equation}
H_{q}=\sum_{\left\{  ij\right\}  }\left( 1-\delta_{\sigma
,\sigma_{j}}\right)  \label{potts}%
\end{equation}
defined on a $d$-dimensional hypercubic lattice $N=L_{1}\times L_{2}%
\times\ldots L_{d}$ with $L_{i}=L$ and periodic boundary conditions.
Here, the sum is only over pairs of nearest-neighbor lattice sites
and $\sigma _{i}=\left[  1,2,\ldots q\right]  $ is the spin state at
the \textit{i-th}
site. This model system is just a generalization of the known Ising model:%
\begin{equation}
H_{\text{Ising}}=-\sum_{\left\{  ij\right\}  }s_{i}s_{j},
\end{equation}
with $s_{i}=\pm1$, which appears as a particular case with $q=2$
after considering a linear transformation among their respective
energy per particle
$\varepsilon_{\text{Ising}}=-d+2\varepsilon_{\text{2-Potts}}$ and
inverse temperature
$\beta_{\text{Ising}}=\frac{1}{2}\beta_{\text{2-Potts}}$.

As discussed elsewhere \cite{mc3}, this model system undergoes a
first-order phase transition when $q>4$ for $d=2$ and $q\geq3$ for
$d=3$, and hence, the microcanonical caloric curves $\beta\left(
\varepsilon\right)$ for these model realizations exhibit the
backbending behavior associated with the presence of macrostates
with $C<0$ such as the one sketched in FIG.\ref{caricature.eps}. In
addition to the consideration of a local MC method as the
\textit{Metropolis importance sample} \cite{metro,Hastings}, the
canonical MC study of Potts models can be carried out by using
another accelerating methods such as the Swendsen-Wang
\cite{SW,pottsm} or Wolff \cite{wolf} \textit{clusters algorithms}.
However, none of these MC algorithms are able to account for the
existence of an anomalous regime with $C<0$, and even, they also
suffer from the existence of a super-critical slowing down as a
direct consequence of the bimodality of the canonical energy
distribution function. The existence of several canonical MC
algorithms for this kind of models enables to perform a comparative
study among their respective extended versions described below.

\subsection{Monte Carlo methods}

\subsubsection{Metropolis importance sample}

The simplest and most general way to implement a thermal coupling
with a bath at constant inverse temperature $\beta_{B}$ is by using
the \textit{Metropolis importance sample} \cite{metro,Hastings}. In
this method, a Metropolis move is carried out as follows:

\begin{enumerate}
\item A site $i$ is chosen at random, and the initial spin state $\sigma_{i}$
is changed (also at random) by considering any other of its $q$
admissible values.

\item This move, from an initial state with energy $E$ and variation $\delta
E$, is accepted in accordance with the transition probability:%
\begin{equation}
W\left(  E\rightarrow E+\Delta E;\beta_{B}\right)  =\min\left\{
\exp\left[ -\beta_{B}\delta E\right]  ,1\right\}  . \label{met.p}
\end{equation}

\end{enumerate}
A MC step is produced after considering $L^{d}$ moves regardless of
whether they have been accepted or rejected.

The extension of this local algorithm with the consideration of an
environment associated with an arbitrary probabilistic weight
$\omega(E)$ is achieved by using the transition probability
(\ref{T1}). Clearly, Eq.(\ref{T1}) fulfils the detailed balance
condition (\ref{DBal}). When the system size $N$ is sufficiently
large, such a transition probability is practically given by
Eq.(\ref{T2}), which is exactly the canonical transition probability
(\ref{met.p}) modified by the inclusion of a variable inverse
temperature, $\beta_{B}\rightarrow\beta_{\omega}\left( E\right)$.
Remarkably, the transition probability (\ref{T1}) can be exactly
rewritten in the form (\ref{detailed.bal.omega}) by using the value
$\beta^{\star}_{\omega}=\beta_{\omega}(E_{m})$ corresponding to
$\theta=0$ for the particular case of the linear ansatz
(\ref{ansatz.linear}). This is a useful representation because both
the initial and the final system configurations, $X$ and $X'$, are
\textit{a priori} known for this local MC method.

\subsubsection{Cluster algorithm}

Other important MC methods are the so-called \textit{cluster
algorithms}, which are usually more efficient than any local MC
method (Metropolis). However, the success of these methods is not
universal because the proper cluster moves needed seem to be highly
dependent on the system, and efficient cluster MC methods have only
been found for a small number of models
\cite{SW,pottsm,wolf,Edwards,Niedermayer,Evertz,Hasenbusch,Dress,Liu}.

The idea of using nonlocal moves was first suggested by
Swendsen-Wang \cite{SW,pottsm} for the case of Ising model and its
generalizations, the Potts models. Such cluster algorithms are based
on a mapping of this model system to a random cluster model of
percolation throughout the equation:
\begin{equation}
Z\left(  \beta_{B}\right)  =\sum_{\left\{  \sigma\right\}
}\exp\left[ -\beta_{B}H_{q}\right]  =\sum_{\left\{  n\right\}
}q^{N_{c}}p^{b}\left(
1-p\right)  ^{Nd-b}, \label{cluster}%
\end{equation}
where $p=p\left(  \beta_{B}\right)  =1-e^{-\beta_{B}}$, $b$ is the
number of bonds and $N_{c}$ the number of clusters. We shall
consider in the present
study the Swendsen-Wang cluster algorithm, whose scheme reads as follows:%

\begin{enumerate}
\item Examine each nearest neighbor pair and create a bond with probability
$p\left(  \beta_{B}\right)  $. That is, if the two nearest neighbor
spins are the same, a bond is created between them with probability
$p\left(  \beta _{B}\right)  $; if spin values are different, there
will be no bond.

\item Identify clusters as a set of sites connected by zero or more bonds
(i.e., connected component of a graph). Relabel each cluster with a
fresh new value at random.
\end{enumerate}

The extension of this cluster MC method by using the present ideas
is achieved by introducing a third step:

\begin{enumerate}
\item[3.] Redefine the inverse temperature of the bath $\beta_{B}%
=\beta_{\omega}\left(  E_{i}\right)$ employed to obtain the next
system configuration $X_{i+1}$ by using the energy $E_{i}=E(X_{i})$
of the present configuration $X_{i}$.
\end{enumerate}

While the bath inverse temperature $\beta_{B}$ is redefined after
every local move in the Metropolis importance sample, such a
redefinition only takes place in a clusters algorithm as the
Swendsen-Wang after every MC step because the clusters moves demand
a constancy of the bath temperature. The present method is a simple
example of extended canonical MC algorithm that does not fulfil the
detailed balance condition (\ref{DBal}) due to its using the
approximated inverse temperature
$\beta^{i}_{\omega}=\beta_{\omega}(E_{i})$ instead of the exact
value $\beta^{\star}_{\omega}=\beta_{\omega}(E^{m}_{i})$ with
$E^{m}_{i}=\left[E_{i}+E_{i+1}\right]/2$. As naturally expected, the
violation of the detailed balance condition introduces finite size
effects in the calculation of the expectation values of the thermal
fluctuations, which affects the calculation of the heat capacity $C$
via Eq.(\ref{cap.fluct}). We have verified by mean of preliminary
calculations that the use of the transition inverse temperature
$\beta^{\star}_{\omega}(E^{m}_{i-1})$ instead of the instantaneous
value $\beta_{\omega}(E_{i})$ in the MC estimation of the
expectation values:
\begin{equation}
\left\langle\beta_{\omega}(E)A(E)\right\rangle \sim\frac{1}{M}\sum^{M}_{i=1}\beta_{\omega}(E_{i})A(E_{i})%
\longrightarrow\frac{1}{M}\sum^{M}_{i=1}\beta^{\star}_{\omega}(E^{m}_{i-1})A(E_{i})
\end{equation}
significantly reduces the incidence of such undesirable errors.

\begin{figure}
[t]
\begin{center}
\includegraphics[
height=2.4in, width=3.5in
]%
{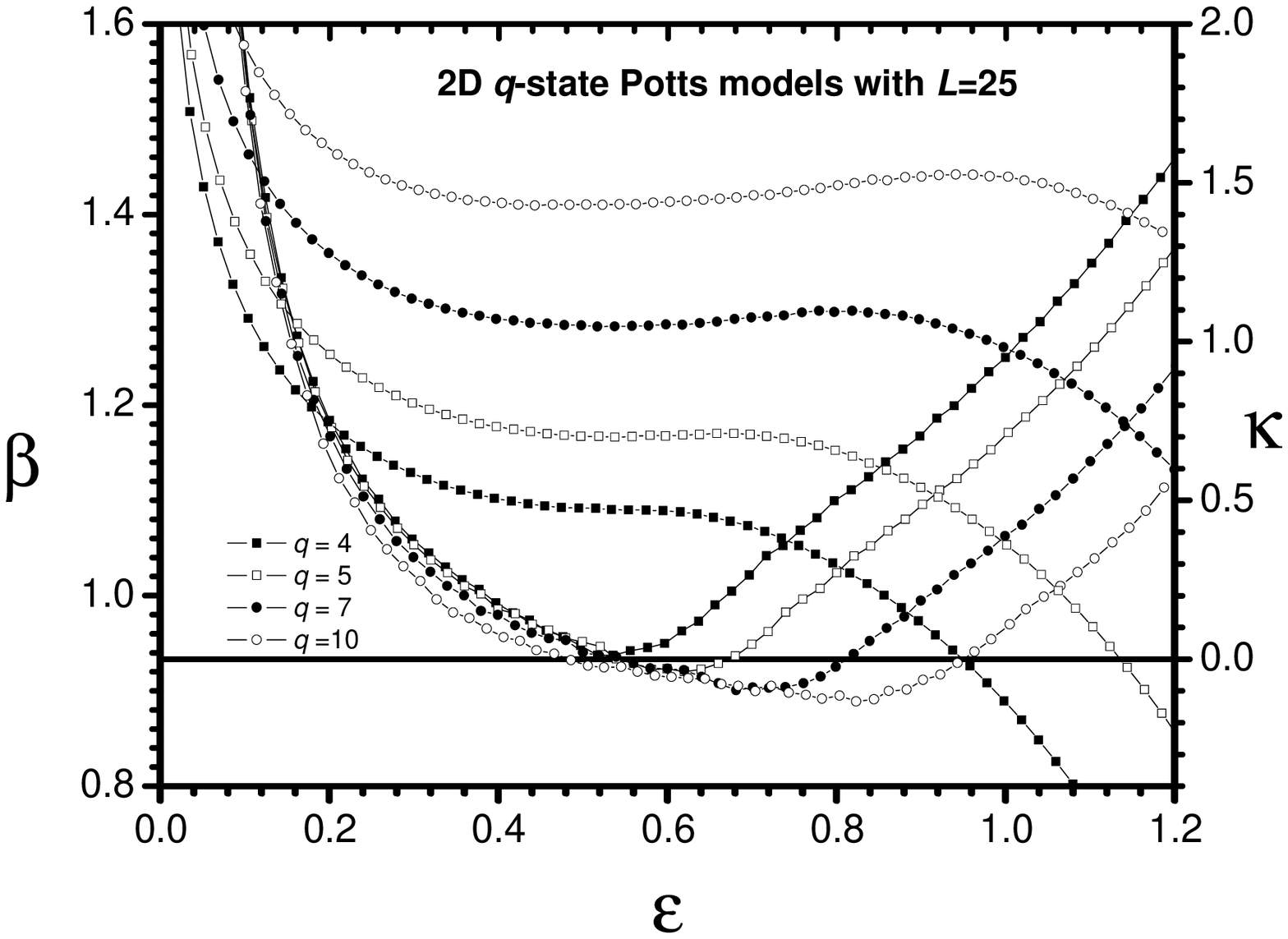}%
\caption{Microcanonical caloric and curvature curves, $\beta\left(
\varepsilon\right)$ (backbending dependencies) and $\kappa\left(
\varepsilon\right)$ (\textit{U-}like dependencies), obtained from MC
simulations by using the extended version of Metropolis importance
sample algorithm for some realizations of 2D $q$-state Potts
models.}
\label{2Dpotts.eps}%
\end{center}
\end{figure}

\subsection{Results and discussions}%

Results of extensive MC calculations by using the extended version
of Metropolis importance sample (hereafter referred to as extended
MIS) are shown in FIG.\ref{2Dpotts.eps}. We are limited to consider
here the 2D Potts models with $L=25$ for several values of $q$,
where each point of these curves has been obtained from a MC run
with $10^{6}$ steps. While the microcanonical curvature curve
$\kappa\left( \varepsilon\right) $ for the case $q=4$ only slightly
touches the horizontal line with $\kappa=0$, the other cases with
$q>4$ clearly exhibit negative values $\kappa\left(
\varepsilon\right) <0$, that is, macrostates with negative heat
capacities $C<0$. This last observation evidences that the 2D Potts
model with $q=4$ undergoes a continuous phase transition, while the
phase transition is discontinuous (first-order) for those cases with
$q>4$ \cite{Wu}.

\begin{figure}
[t]
\begin{center}
\includegraphics[
height=2.4in, width=3.5in
]%
{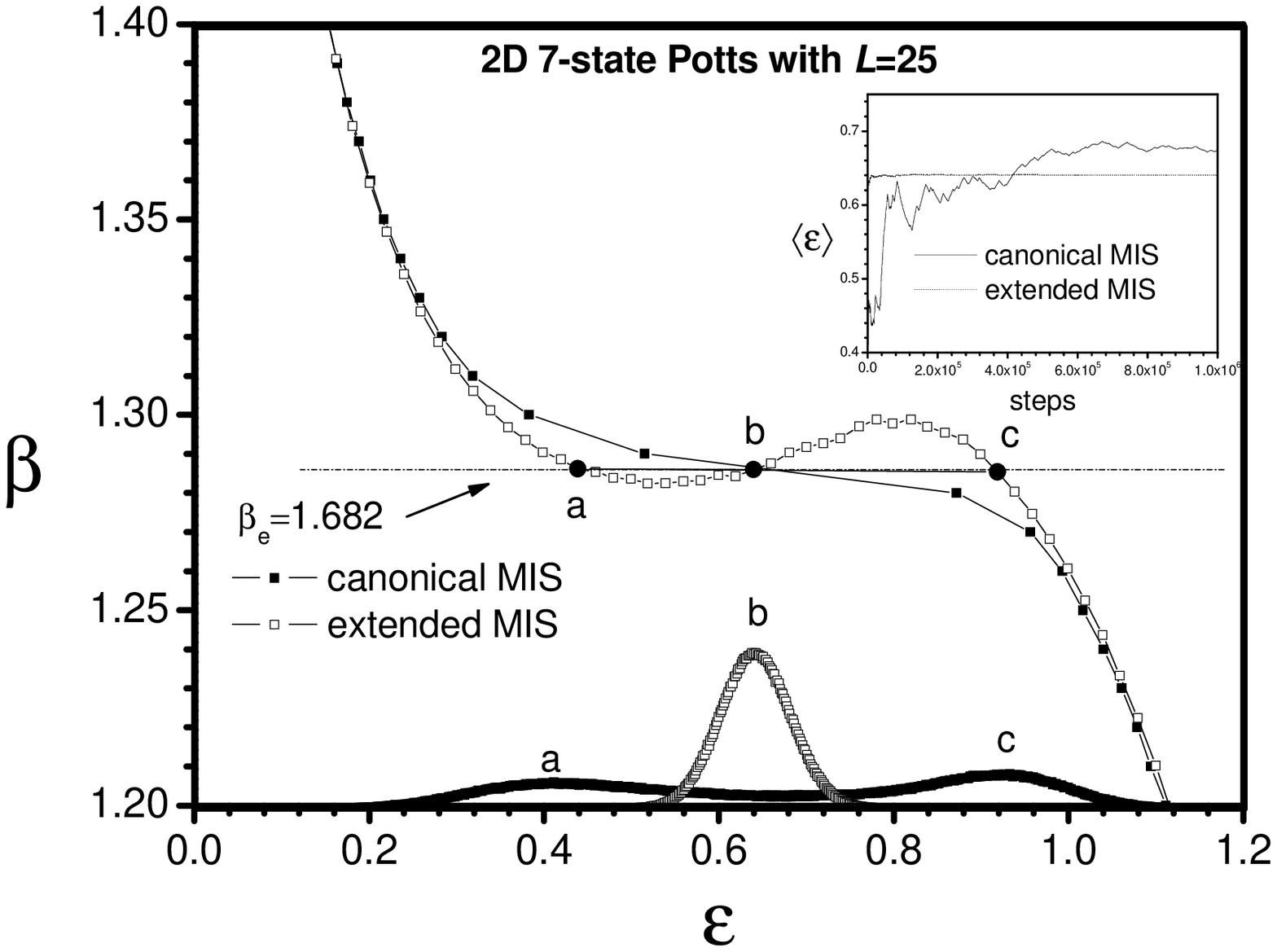}%
\caption{Comparison between the caloric curves of the 2D $7$-state
Potts model with $L=25$ obtained from MC simulations by using the
canonical Metropolis importance sample (canonical MIS) and its
extended version (extended MIS). The distribution functions as well
as the evolution of the average system energy $\left\langle
\varepsilon\right\rangle $ shown in the inset panel correspond to MC
runs by using these methods with $\beta_{e}=1.682$. See the text for
further explanations.}
\label{comparative.eps}%
\end{center}
\end{figure}

As expected, the anomalous macrostates with $C<0$ cannot be accessed
by using the ordinary Metropolis importance sample (hereafter
referred to as canonical MIS). This limitation is explicitly shown
in FIG.\ref{comparative.eps} for the particular case of the 2D
$7$-state Potts model. These results constitute a simple
exemplification of the schematic behavior represented in
FIG.\ref{caricature.eps}. For a better understanding, we also show
here the corresponding energy distribution functions and the
dynamical evolutions of the average system energy $\left\langle
\varepsilon\right\rangle $ (inset panel) obtained from MC runs by
considering both canonical MIS and extended MIS algorithms with
$\beta_{e}=1.682$.

The stationary macrostates at $\varepsilon_{a}$ and
$\varepsilon_{c}$ derived from the intersection of the
microcanonical curve $\beta\left(  \varepsilon \right)$ and the bath
inverse temperature $\beta_{B}=\beta_{e}=1.682$ are
thermodynamically stable within the canonical ensemble.
Consequently, the canonical energy distribution function is bimodal
and its peaks are related to the coexisting phases. As consequence
of such bimodality, the system energy exhibits eventual random
transitions between the coexisting peaks, which lead to a slow
equilibration of the corresponding average energy $\left\langle
\varepsilon\right\rangle $ (inset panel).

The thermodynamic behavior radically changes when this system is put
in a thermodynamic situation with non-vanishing correlated
fluctuations $\left\langle \delta\beta_{\omega}\delta E\right\rangle
\not =0$, which is implemented here by considering a dependence
$\beta_{\omega}(\varepsilon)=\beta_{e}+\lambda(\varepsilon-\varepsilon_{e})$,
where $\beta_{e}=1.682$,
$\lambda=\lambda_{\text{opt}}(\kappa_{b})\not =0$ and
$\varepsilon_{e}=\varepsilon_{b}$, with $\kappa_{b}$ being the
curvature at the stationary point $\varepsilon_{b}$ located within
the anomalous region with $C<0$. The canonically stable stationary
points $\varepsilon_{a}$ and $\varepsilon_{c}$ become
thermodynamically unstable and their corresponding peaks disappear
from the energy distribution function. Conversely, the canonically
unstable stationary point $\varepsilon_{b}$ now becomes
thermodynamically stable, and its position determines the maximum of
the only peak of the energy distribution function. Once the bimodal
character of the energy distribution function is suppressed, the
average system energy $\left\langle \varepsilon\right\rangle $ shows
fast convergence towards its equilibrium value $\varepsilon_{b}$
(inset panel).

\begin{figure}
[t]
\begin{center}
\includegraphics[
height=2.4in, width=3.5in
]%
{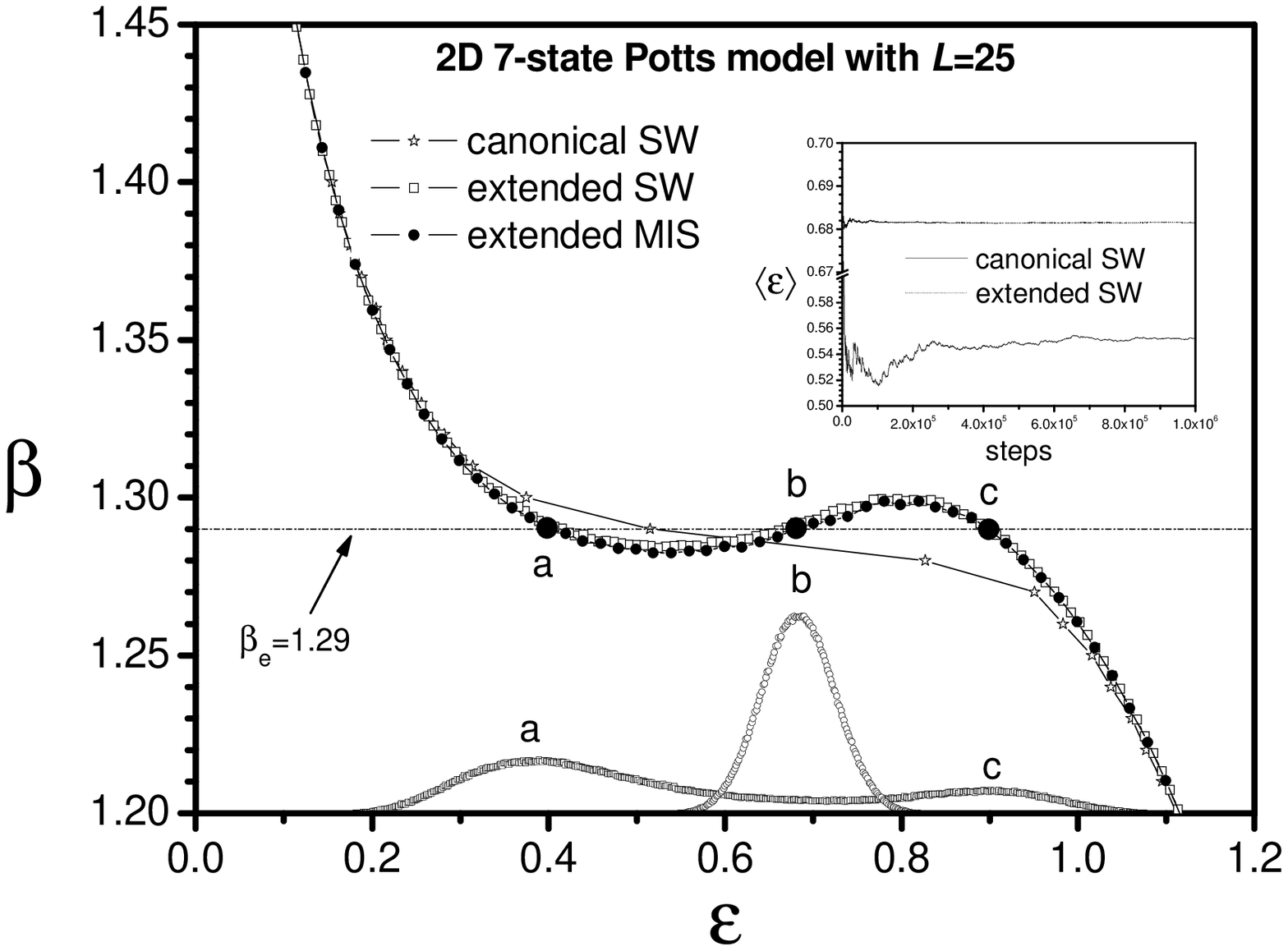}%
\caption{Comparison among the caloric curves of the 2D $7$-state
Potts model with $L=25$ obtained from MC simulations using the
canonical Swendsen-Wang clusters algorithm (canonical SW) and its
extended version (extended SW), as well as the extended Metropolis
importance sample method (extended MIS). The energy distribution
functions and the dynamical evolution of the average system energy
$\left\langle \varepsilon\right\rangle $ shown in the inset panel
correspond to MC runs by using the above clusters algorithms with
$\beta_{e}=1.29$.}
\label{swcase.eps}%
\end{center}
\end{figure}

A comparative study of a 2D 7-state Potts model with $L=25$ by using
the Swendsen-Wang cluster algorithm is shown in
FIG.\ref{swcase.eps}. Although the canonical Swendsen-Wang method
(hereafter referred to as canonical SW) is more efficient than the
canonical MIS, it is unable to describe the existence of an
anomalous region with $C<0$ and suffers from a slow relaxation in
this kind of situation, as it is clearly shown in
FIG.\ref{swcase.eps}. Such limitations are circumvented by using its
extended version (hereafter referred to as extended SW). As the
previously discussed example, the extended SW algorithm eliminates
the bimodality of the energy distribution function for
$\beta_{e}=1.29$ and leads to a fast convergence of the average
energy per particle along the simulation. Note also the remarkable
agreement between the extended MIS and extended SW algorithms
despite the latter one does not fulfil the detailed balance
condition (\ref{DBal}). This is a clear evidence that a suitable
approximation of the inverse temperature $\beta^{\star}_{\omega}$ of
Eq.(\ref{detailed.bal.omega}) is enough to provide a precise
estimation of the microcanonical caloric curve $\beta(E)$ as long as
the system size is sufficiently large.

\begin{figure}
[t]
\begin{center}
\includegraphics[
height=2.4in, width=3.5in
]%
{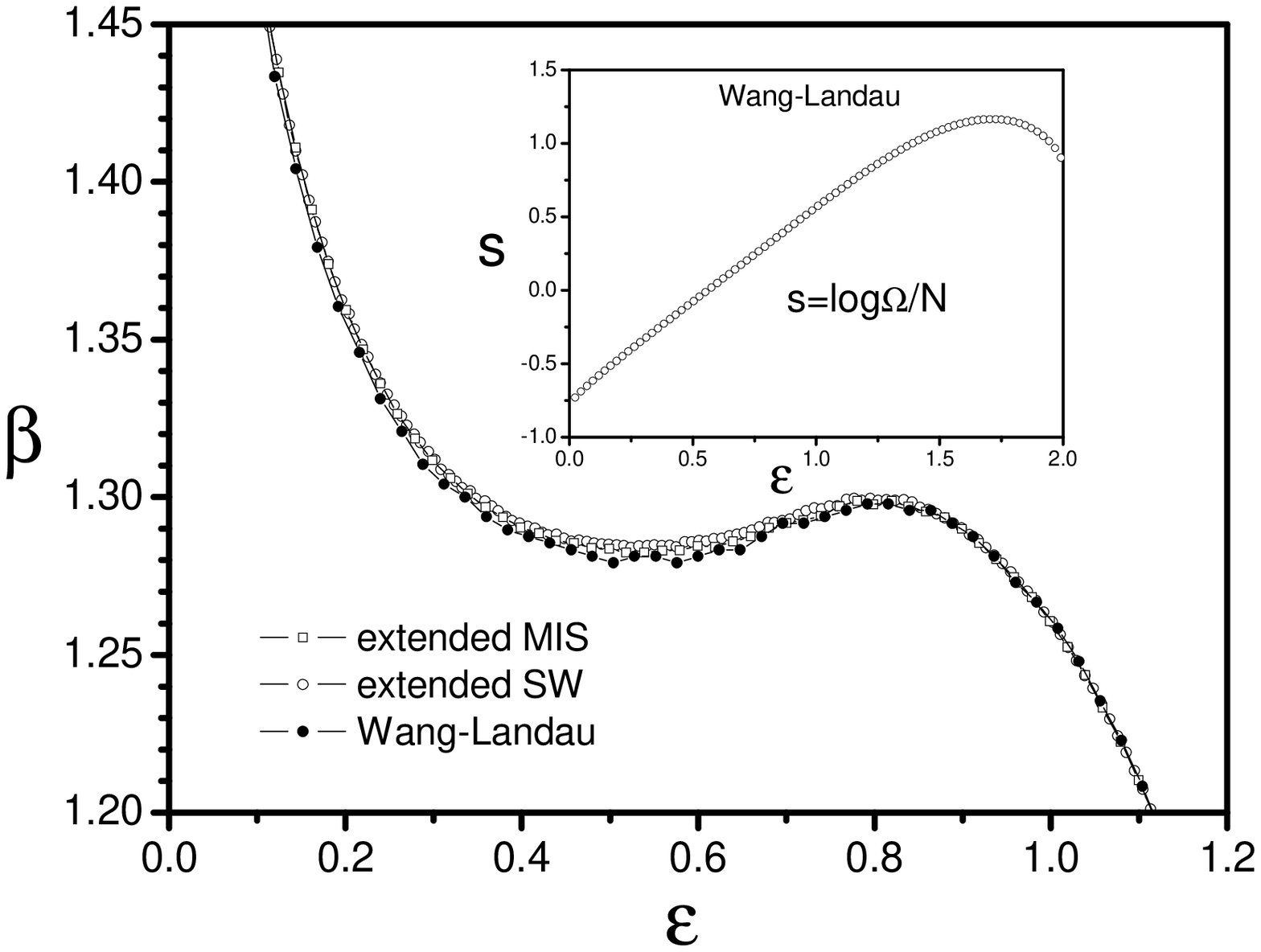}%
\caption{Comparative study between the extended MIS and SW
algorithms with the known Wang-Landau sampling method. Despite of
the relative good agreement, one can appreciate some small
discrepancies, which could be mainly associated with size effects.}
\label{compWLSW.eps}
\end{center}
\end{figure}

The comparison among the above extended canonical MC algorithms and
the known Wang-Landau sampling method \cite{WangLandau} is shown in
FIG.\ref{compWLSW.eps}. Although these results exhibit very good
consistency, one can note small but appreciable discrepancies within
the anomalous region with $C<0$. These relative differences are
naturally expected due to two reasons. Firstly, a direct numerical
differentiation of the entropy $S=\log\Omega$ (inset panel) obtained
from the Wang-Landau method enhances the underlying statistical
errors of its MC calculation, an therefore, the final result
crucially depends on how one defines this mathematical operation for
this discrete observable. On the other hand, the work equation
(\ref{ave.inter}) is supported by the Gaussian character of the
energy distribution function $\rho(\varepsilon)$, which arises as an
asymptotic distribution when the system size $N$ is sufficiently
large. Clearly, the distribution function undergoes small deviations
from the Gaussian shape when $N$ is not as large. We shall show in a
forthcoming paper that these size effects can be taken into account
to improve the precision of some extended canonical MC algorithms.

\begin{figure}
[t]
\begin{center}
\includegraphics[
height=2.4in, width=3.5in
]%
{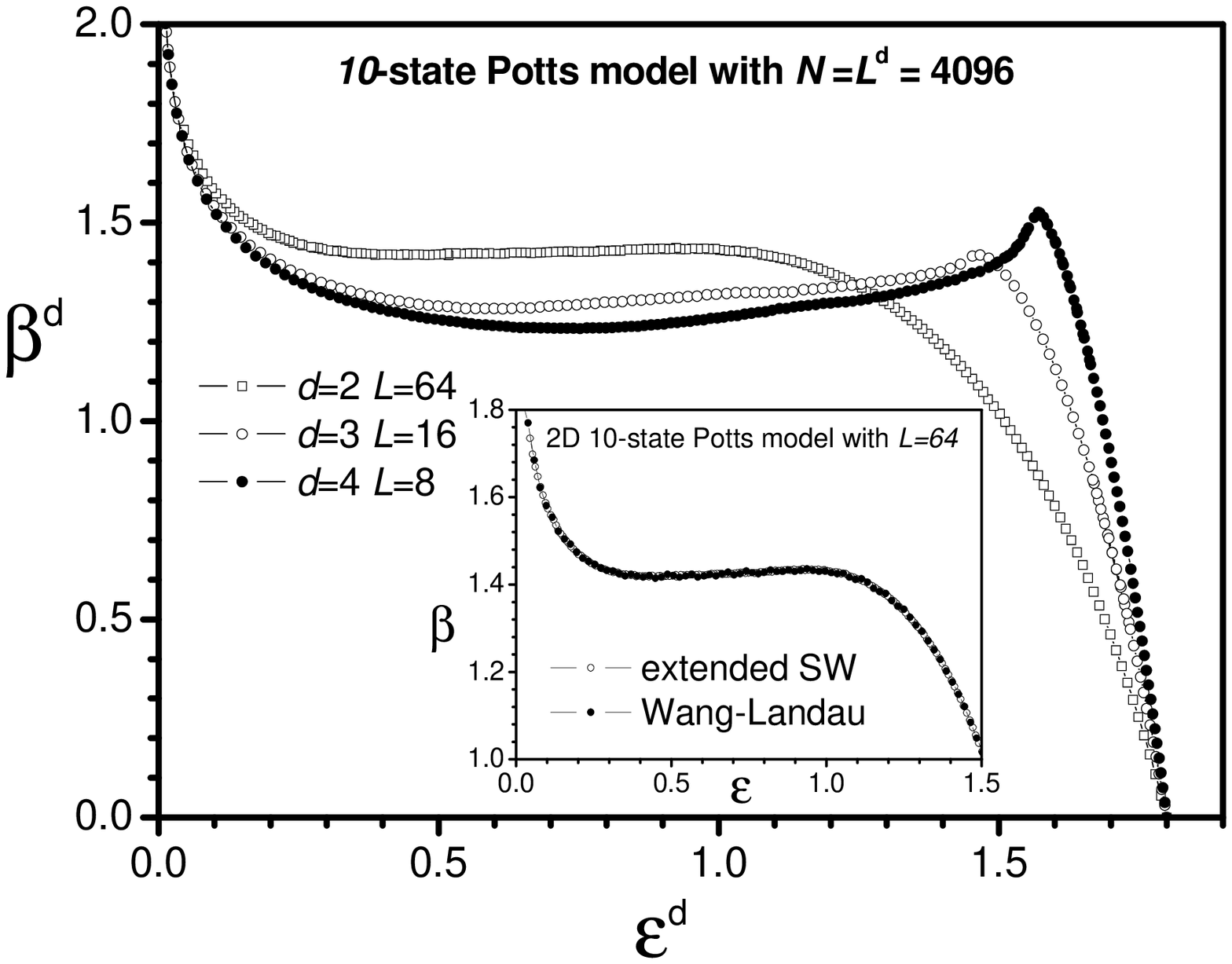}%
\caption{Microcanonical caloric curves of the $10$-state Potts
models for lattice dimensions $d=\left(  2,3,4\right)  $ with a
fixed number of lattice sites $N=L^{d}=4096$, which were obtained
from MC simulations by using the extended SW algorithm with
$2\times10^{4}$ MC steps for each calculated point. The plot is
performed in terms of the re-scaled variables
$\varepsilon^{d}=2\varepsilon/d$ and $\beta^{d}=\beta d/2$. Inset:
Comparison between extended SW and Wang-Landau method for the 2D
$10$-state Potts model with $L=64$.}
\label{caloric.dim.eps}%
\end{center}
\end{figure}

Overall, it is important to remark that the present proposal is
focussed on the solution of the slow sampling problems observed in
large-scale canonical MC simulations, that is, in systems undergoing
a temperature driven discontinuous phase transition with sizes
sufficiently large to support the Gaussian approximation. In
particular, the previous extended canonical MC algorithms can be
useful to study Potts models with many spin states and higher
dimensions, as the cases shown in FIG.\ref{caloric.dim.eps}, where
we illustrate the caloric curves derived from MC simulations by
using the extended SW algorithm for $q=10$, $d=\left( 2,3,4\right)$,
and a fixed number of lattice sites $N=L^{d}=4096$. The comparative
study case with the Wang-Landau method shown in the inset panel
allows us to verify that the agreement between these methods is more
significant with the increasing of $N$.

To quantitatively characterize the efficiency of the previously
discussed extended canonical MC methods, one should obtain the
\textit{decorrelation time} $\tau$, that is, the minimum number of
MC steps needed to generate effectively independent, identically
distributed samples in the Markov chain. Its calculation is
performed here by using the expression:
\begin{equation}
\tau=\lim_{M\rightarrow\infty}\tau_{M}=\lim_{M\rightarrow\infty}\frac{M\cdot
var\left(  \varepsilon_{M}\right)  }{var\left(
\varepsilon_{1}\right)  },
\end{equation}
where $var\left(  \varepsilon_{M}\right)  =\left\langle \varepsilon_{M}%
^{2}\right\rangle -\left\langle \varepsilon_{M}\right\rangle ^{2}$
is the variance of $\varepsilon_{M}$, which is defined as the
arithmetic mean of the
energy per particle $\varepsilon$ over $M$ samples (consecutive MC steps):%
\begin{equation}
\varepsilon_{M}=\frac{1}{M}\sum_{i=1}^{M}\varepsilon_{i}.
\end{equation}
This quantity is calculated for the particular MC runs considered in
FIG.\ref{comparative.eps} and FIG.\ref{swcase.eps}. This study
allows us to verify that the use of a variable dependence
$\beta_{\omega}(\varepsilon)$ instead of a constant parameter
$\beta_{B}$ enables the reduction of the decorrelation time of the
Metropolis importance sample from $\tau\simeq450$ to $\tau\simeq80$.
The improvement is even more significant for the Swendsen-Wang
clusters algorithm, which experiences a reduction of the
decorrelation time from $\tau\simeq300$ to $\tau\simeq12.2$.

\begin{figure}
[t]
\begin{center}
\includegraphics[
height=4.4in, width=3.4in
]%
{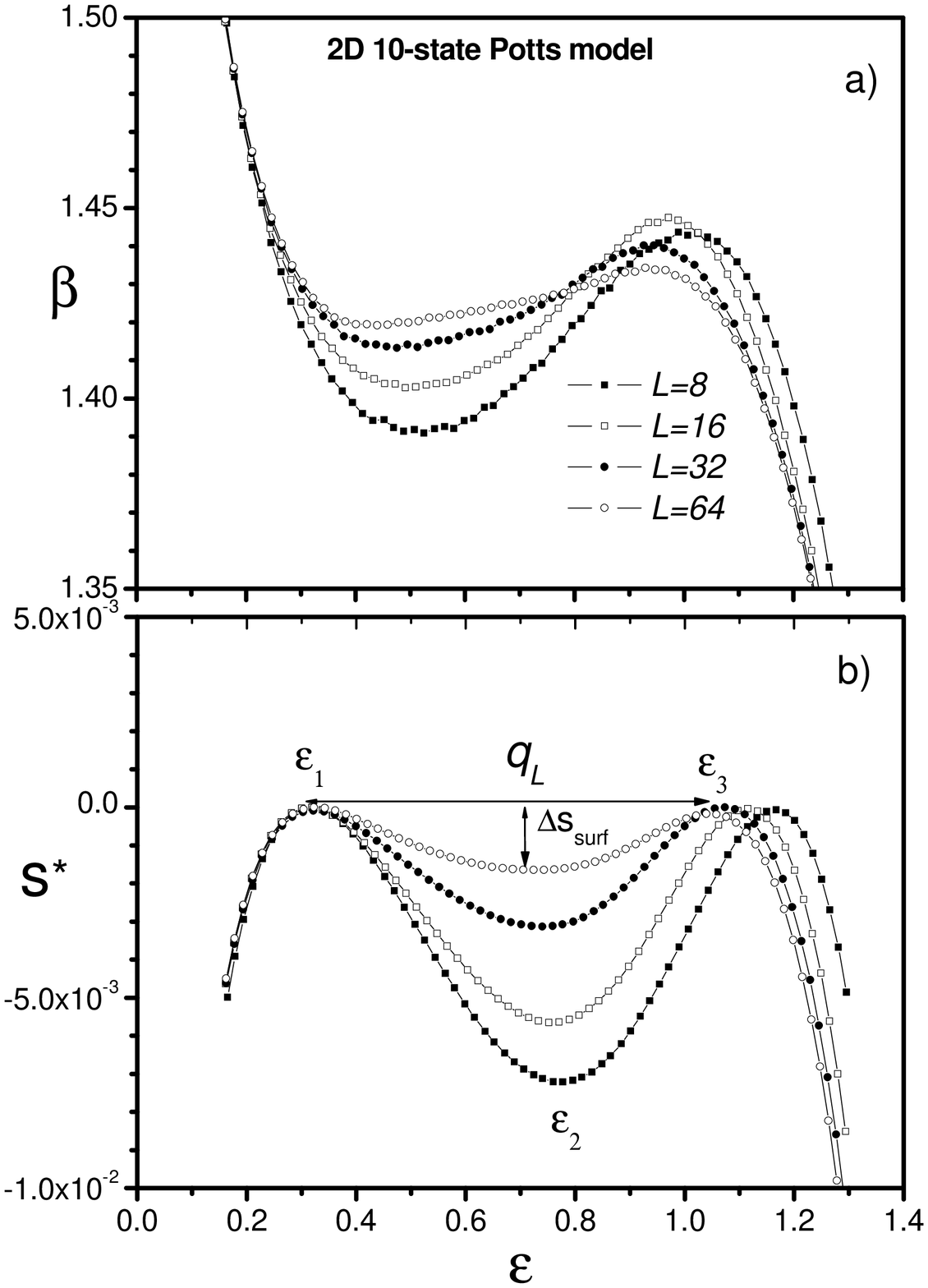}%
\caption{Study of the thermodynamic behavior of the 2D $10$-state
Potts model for different lattice sizes $L$: Panel a) microcanonical
caloric curves $\beta\left(  \varepsilon\right)  $, Panel b)
microcanonical entropy plotted in term of the quantity
$s^{\ast}\left(  \varepsilon\right)  \equiv s\left(
\varepsilon\right)  -\beta_{c}\varepsilon-s_{0}$ in order to reveal
the \textit{convex intruder} associated with the existence of
surface correlations.}
\label{size.eps}%
\end{center}
\end{figure}

\begin{figure}
[t]
\begin{center}
\includegraphics[
height=2.4in, width=3.5in
]%
{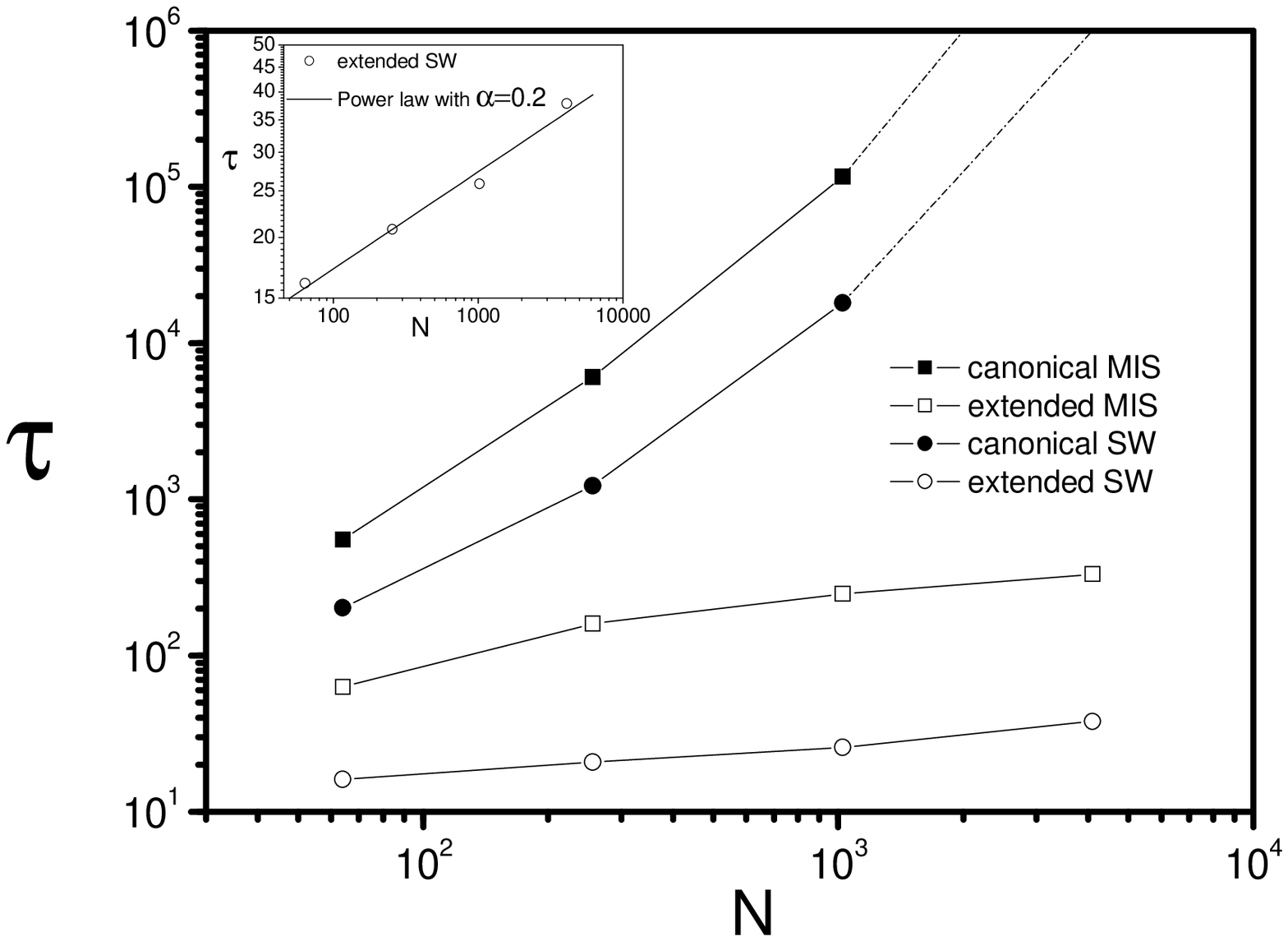}%
\caption{Dependencies of the decorrelation time $\tau$
\textit{versus} the system size $N$ at the point of first-order
phase transition $\beta_{c}$ for different MC methods considered in
the study of the $2D$ $10$-state Potts model. Inset panel: A more
detailed plot for the decorrelation time of the extended SW method,
which clearly shows a weak power-law dependence $\tau\propto
N^{\alpha}$ with
$\alpha=0.2$.}%
\label{decorrelation.eps}%
\end{center}
\end{figure}

Let us now analyze the behavior of the decorrelation time $\tau$
with the increasing system size $N=L^{d}$ at the inverse temperature
$\beta_{c}$ of the first-order phase transition. Such a study
demands the performance of preliminary calculations of the quantity
$\beta_{c}$ due to its underlying dependence on the system size $N$.
For computational limitations, we decide to restrict our analysis
for the case of 2D 10-state Potts model with $L=\left(
8,16,32,64\right) $, whose results are shown in FIG.\ref{size.eps}.
The entropy per particle $s\left( \varepsilon\right) $ was obtained
from the second-order approximation of
the power-expansion $\Delta s=\beta\Delta\varepsilon-\frac{1}{2}%
\kappa\Delta\varepsilon^{2}$, which allows us a direct calculation
of this thermodynamic function through the inverse temperature
$\beta$ and curvature $\kappa$. These numerical results are shown in
panel b) of FIG.\ref{size.eps}, or more exactly, the quantity
$s^{\ast}\left( \varepsilon\right) =s\left(  \varepsilon \right)
-\beta_{c}\varepsilon-s_{0}$, with $s_{0}$ being a suitable
constant, which allows us to appreciate better the \textit{convex
intruder} of the entropy accounting for the existence of macrostates
with $C<0$. The mathematical form of this last anomaly enables the
acquisition of the latent heat
$q_{L}=\varepsilon_{3}-\varepsilon_{1}$ and the entropy-loss per
particle $\Delta s_{surf}$ associated with the existence of
\textit{surface correlations} \cite{gro1}, and $\left(
\varepsilon_{1},\varepsilon _{2},\varepsilon_{3}\right)  $ represent
the three stationary points where the caloric curve $\beta\left(
\varepsilon\right)  $ takes the value of the phase transition
inverse temperature $\beta_{c}$. It worth to clarify that the
critical value $\beta_{c}$ is simply the point of discontinuity of
the first derivative of the Planck thermodynamic potential
$p(\beta)$ estimated from the microcanonical entropy
$s(\varepsilon)$ via the Legrendre's transformation:
\begin{equation}
p(\beta)=\min_{E}\left[\beta \varepsilon-s(\varepsilon)\right].
\end{equation}
In fact, once obtained the microcanonical entropy $S(E)$, one can
calculate any thermo-statistical quantity by using the canonical
distribution function:
\begin{equation}
dp\left( E\left\vert \beta \right. \right) =Ae^{-\beta E+S\left(
E\right) }dE.
\end{equation}
Relevant physical observables and thermodynamic parameters derived
from this type of analysis are summarized in Table
\ref{observables}.

\begin{table}[tbp] \centering
\begin{tabular}
[c]{|c|c|c|c|c|}\hline\hline $L$ & $8$ & $16$ & $32$ &
$64$\\\hline\hline $\beta_{c}$ & $1.415$ & $1.422$ & $1.424$ &
$1.426$\\\hline $\varepsilon_{1}$ & $0.319$ & $0.319$ & $0.321$ &
$0.329$\\\hline $\varepsilon_{2}$ & $0.767$ & $0.755$ & $0.737$ &
$0.719$\\\hline $\varepsilon_{3}$ & $1.165$ & $1.114$ & $1.074$ &
$1.049$\\\hline $q_{L}$ & $0.\,\allowbreak846$ & $0.795\,$ &
$0.753\,$ & $\allowbreak 0.72$\\\hline $\Delta s_{surf}\times10^{3}$
& $72$ & $5.6$ & $3.1$ & $1.7$\\\hline\hline
\end{tabular}
\caption{Dependence of some physical observables and thermodynamic
parameters of the 2D $10$-states Potts model with the increasing
lattice size $L$.}\label{observables}
\end{table}%

Results of extensive calculations of dependencies of the
decorrelation time $\tau$ \textit{versus} the system size $N$ at the
point of phase transition $\beta_{c}$ are shown in
FIG.\ref{decorrelation.eps}. It is clearly evident that the extended
MC methods are always much more efficient than their respective
canonical counterparts. In particular, the extended SW method
reduces the exponential growth of the decorrelation time $\tau$ with
$N$ to a weak power-law dependence $\tau\propto N^{\alpha}$ with
$\alpha=0.2$. As expected, the extended MIS is less efficient than
the extended SW. However, the $N$-dependence of its corresponding
decorrelation time $\tau$ does not differ in a significant way from
the one associated with the cluster algorithm. Since the calculation
of the decorrelation time $\tau$ of the canonical MC methods demands
very large MC runs, we use the $1/M$ extrapolation in order to
obtain some rough estimates of the non-equilibrated points.

According to Berg \cite{Berg97}, the multicanonical method and its
variant are able to reduce the exponential divergence of the
decorrelation time $\tau$ to a power dependence with typical
exponent $\alpha=2-2.5$. Although such a power-law behavior accounts
for a less efficient convergence than the one achieved by using
present proposal, it is important to remark that the multicanonical
methods possibilities an effective exploration of the entire energy
region in a single MC run. Conversely, our methodology only explores
a small energy region
$(\varepsilon_{e}-\sigma,\varepsilon_{e}+\sigma)$ with a typical
width $\sigma\sim1/\sqrt{N}$, so that, a minimum of
$n\propto\sqrt{N}$ MC runs are required to study the same energy
range of multicanonical methods. Thus, the size dependence the
number of MC samples $M$ can be estimated by considering the number
of MC runs $n$ and the decorrelation time $\tau$ as follows:
\begin{equation}
M\sim n\tau\sim N^{\alpha^{*}},
\end{equation}
which leads to the effective exponent $\alpha^{*}\simeq0.7$. Such an
effective power-law growth still considers a more efficient
convergence rate than the one associated with multicanonical
methods, overall, when one also takes into account the fact that
these re-weighting techniques demand a preliminary reconstruction of
the probabilistic weight $\omega(E)$, which is a procedure that
consumes a significant fraction of the available computational
resources.

\begin{figure}
[t]
\begin{center}
\includegraphics[
height=2.4in, width=3.4in
]%
{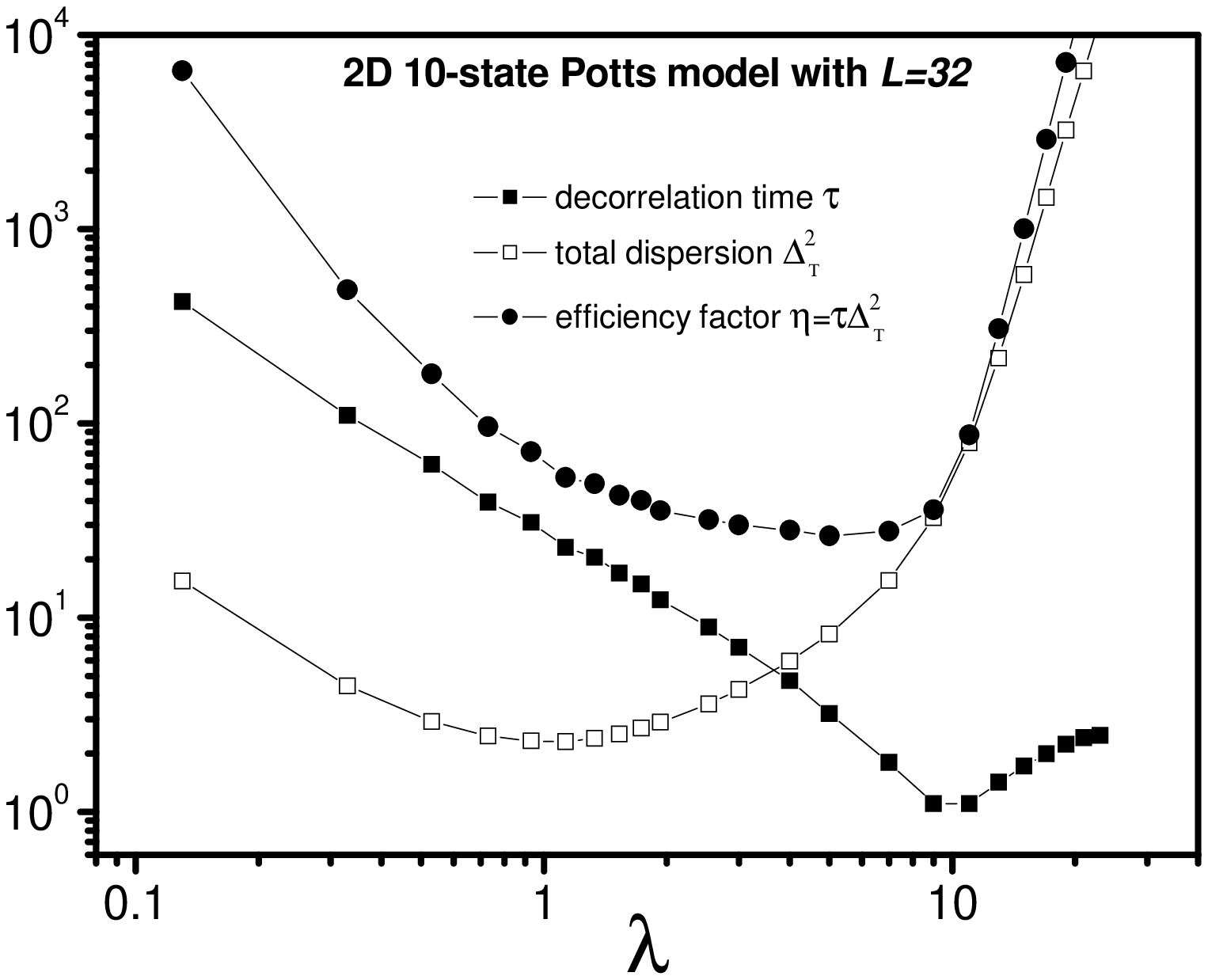}%
\caption{Dependence of the decorrelation time $\tau$, the total
dispersion $\Delta^{2}_{T}$ and the efficiency factor
$\eta=\tau\Delta^{2}_{T}$ on the coupling constant $\lambda$
obtained from MC simulations of the 2D $10$-state Potts model with
$L=32$ by using the extended SW clusters algorithm. See the text for
further explanations.}
\label{lambda.eps}%
\end{center}
\end{figure}

Let us finally reconsider the question about the optimal value of
the coupling constant $\lambda$ that allows to achieve the best
efficiency by using an extended canonical MC algorithm. An example
of such a study is shown in FIG.\ref{lambda.eps}, where we
illustrate the behavior of the decorrelation time $\tau$ and the
total dispersion $\Delta^{2}_{T}=(\Delta E)^{2}/N+N(\Delta
\beta_{\omega})^{2}$ for positive values of the coupling constant
$\lambda$. The data was obtained from MC simulations of the 2D
$10$-state Potts model with $L=32$ by using the extended SW clusters
algorithm with environment inverse temperature
$\beta_{\omega}(\varepsilon)=\beta_{e}+\lambda(\varepsilon-\varepsilon_{b})$,
whose parameters $\beta_{e}=\beta_{c}$ and
$\varepsilon_{e}=\varepsilon_{2}$ in order to access the canonically
unstable stationary macrostate $\varepsilon_{2}$ located within the
anomalous region with $C<0$ at the point of the first-order phase
transition $\beta_{c}$.

As expected, the decorrelation time $\tau$ and the total dispersion
$\Delta^{2}_{T}$ show large values when the coupling constant
$\lambda$ is close to zero as a consequence of the imposition of the
external conditions associated with the canonical ensemble
(\ref{Gibbs}), where take place thermodynamic instability of the
macrostates with $C<0$ and the bimodal character of the energy
distribution function $\rho(\varepsilon)$. The increase of the
coupling parameter $\lambda$ for values that obeys the condition
(\ref{CEq}) produces a progressive reduction of the energy
dispersion $\Delta E$ in Eq.(\ref{dispersions}). This last behavior
leads to a reduction of the decorrelation time $\tau$ due to the MC
sampling performs a faster exploration for a smaller energetic
region. However, the decorrelation time $\tau$ starts to grow after
reach its minimum value at a certain $\lambda_{\tau}$ since the
system configurations $X$ cannot be modified in an appreciable way
by the MC sampling if the energy is only allowed to undergo too
small changes around its equilibrium value.

For a general case of the environment inverse temperature
(\ref{ansatz.linear}) with $\lambda\neq0$, both the system
fluctuating behavior and the decorrelation time depend on the
particular value of the coupling constant $\lambda$. In such cases,
the efficiency of an extended canonical MC method can be evaluated
throughout the minimal number of MC steps needed to achieve a
certain precision in the calculation of a given thermo-statistical
observable. The statistical error $\delta\varepsilon$ associated
with a set of $n$ independent outcomes $\{x_{i},i=1,..n\}$ can be
estimated from the standard deviation $\sigma$ as
$\delta\epsilon^{2}=\sigma^{2}/n$. In MC calculations, independent
samples are only obtained after considering a number of MC steps
equal to the decorrelation time $\tau$. Thus, the number of MC steps
$S$ needed to obtain an estimation of a given point
$(\varepsilon,\beta)$ of the caloric curve with a precision
$\sqrt{\delta\varepsilon^{2}+\delta\beta^{2}}<\delta a$ can be
evaluated in terms of the decorrelation time $\tau$, the total
dispersion $\Delta^{2}_{T}$ and the system size $N$ as follows:
\begin{equation}
S=\frac{\tau\Delta^{2}_{T}}{N\delta a^{2}}\equiv\frac{\eta}{N\delta
a^{2}}.
\end{equation}
The quantity $\eta=\tau\Delta^{2}_{T}$ could be referred to as the
\textit{efficiency factor}. Clearly, a \textit{dynamic criterion} in
order to provide an optimal value of the coupling constant $\lambda$
to achieve the best efficiency by using this kind of MC calculations
is given by minimizing the efficiency factor $\eta$.

According to the data shown in FIG.\ref{lambda.eps}, the value of
the coupling constant $\lambda_{\eta}$ corresponding to the minimum
of the efficiency factor $\eta$ is located within the interval
between the position of the minimum $\lambda_{\tau}$ of the
decorrelation time $\tau$ and the minimum $\lambda_{s}$ of the total
dispersion $\Delta^{2}_{T}$,
$\lambda\in\left[\lambda_{s},\lambda_{\tau}\right]$. Remarkably, the
efficiency factor $\eta$ does not change in a significant way within
this last interval. By considering this last observation and the
fact that the calculation of the correlation time $\tau$ demands
extensive calculations, one can realize that the value of the
coupling constant $\lambda_{s}$ corresponding to the minimum of the
total dispersion $\Delta^{2}_{T}$, Eq.(\ref{optimal}), provides a
good value in order to perform a very efficient MC calculation of
the caloric curve by using extended canonical algorithms. In
general, it is always convenient to keep as small as possible the
value of the coupling constant $\lambda$. Indeed, the statistical
error $\epsilon$ involved in the MC calculation of the standard
deviation $\sigma^{2}_{e}\equiv(\Delta E)^{2}/N$ considered to
obtain the system curvature $\kappa=\beta^{2}N/C$ from
Eq.(\ref{cap.fluct}) leads to the existence of a statistical error
$\delta\kappa\sim(\lambda+1)(\lambda+\kappa)\epsilon$, which grows
with the increase of the coupling constant $\lambda$.

\section{Final remarks\label{conclusions}}

We have shown that conventional Monte Carlo methods based on the
consideration of the Gibbs canonical ensemble (\ref{Gibbs}) can be
easily extended in order to capture the existence of an anomalous
regime with negative heat capacities and avoid the incidence of the
super-critical slowing down. The key ingredient is to replace the
use of a bath with an infinite heat capacity by an environment with
a finite heat capacity that obeys Thirring's constrain
(\ref{thir.const}). Such an equilibrium situation, characterized by
the existence of non-vanishing correlations $\left\langle
\delta\beta_{\omega}\delta E\right\rangle \neq0$, is inspired on the
generalized equilibrium fluctuation-dissipation relation
(\ref{tur}), which allows to introduce several improvements to the
methodology of Gerling and H\"{u}ller based on the consideration of
the dynamical ensemble (\ref{GHens}).

The way to introduce an environment with a variable inverse
temperature $\beta_{\omega}$ depends on the own features of each
canonical Monte Carlo method, although such a question seems not to
be a difficult problem in the case of classical algorithms. While it
could be desirable that the implementation of this kind of
methodology obeyed the detailed balance condition (\ref{DBal}), the
application examples considered in the section \ref{examples} show
that one can still obtain a good MC estimation of the microcanonical
caloric curve $\beta(E)$ without fulfilling the detailed balance as
long as the system under analysis be sufficiently large.

Before concluding this section, it is worthwhile to mention that
Eq.(\ref{tur}) constitutes a particular case of a more general
equilibrium fluctuation-dissipation theorem, which accounts for the
system fluctuating behavior in a thermodynamic situation
characterized by the incidence of several control parameters
\cite{vel-gfdt}. Roughly speaking, this theorem provides a general
extension of some other well-known fluctuation relations such as the
one involving the isothermal magnetic susceptibility and
magnetization fluctuations of a magnetic system,
$\chi_{T}=\beta\left\langle\delta M^{2}\right\rangle$, or the
isothermal compressibility and volume fluctuations of a fluid
system, $VK_{T}=\beta\left\langle\delta V^{2}\right\rangle$, which
are compatible with the existence of anomalous response functions,
e.g., negative isothermal susceptibilities $\chi_{T}<0$ or negative
isothermal compressibilities $K_{T}<0$. Clearly, this general
theorem suggests a direct extension of the present methodology in
order to enhance MC methods based on the so-called Boltzmann-Gibbs
distributions:
\begin{equation}\label{BG}
dp_{BG}(E,X)=\frac{1}{Z(\beta,Y)}\exp\left[-\beta(E+YX)\right]dEdX
\end{equation}
to account for the existence of macrostates with anomalous response
functions.

\section*{Acknowledgments}

It is a pleasure to acknowledge partial financial support by
FONDECYT 3080003. L.Velazquez also thanks the partial financial
support by the project PNCB-16/2004 of the Cuban National Programme
of Basic Sciences.


\begin{thebibliography}{99}
\bibitem {pad}T. Padmanabhan, Physics Reports \textbf{188}, 285 (1990).

\bibitem {Lyn3}D. Lynden-Bell, Physica \textbf{A 263, }293 (1999).

\bibitem {gro1}D. H. E. Gross, \textit{Microcanonical thermodynamics: Phase
transitions in Small systems}, \textit{66 Lectures Notes in
Physics}, (World scientific, Singapore 2001).

\bibitem {gro na}D. H. E. Gross and M. E. Madjet, Z. Phys. \textbf{B 104}
(1997) 521.

\bibitem {moretto}L.G. Moretto, R. Ghetti, L. Phair, K. Tso, G.J. Wozniak,
Phys. Rep. \textbf{287} (1997) 250.

\bibitem {Dagostino}M. D'Agostino, F. Gulminelli, P. Chomaz, M. Bruno, F.
Cannata, R. Bougault, F. Gramegna, I. Iori, N. Le Neindre, GV.
Margagliotti, A. Moroni and G. Vannini, Phys. Lett. \textbf{B 473}
(2000) 219.

\bibitem {mc3}P. D. Landau and K. Binder, \textit{A guide to Monte Carlo
simulations in Statistical Physics} (Cambridge Univ Press, 2000).

\bibitem {vel-tur}L. Velazquez and S. Curilef, J. Phys. A: Math. Theor.
\textbf{42}, 095006 (2009).

\bibitem {vel-jstat}L. Velazquez and S. Curilef, J. Stat. Mech. P03027 (2009).

\bibitem {metro}N. Metropolis, A. W. Rosenbluth, M. N. Rosenbluth, A. H.
Teller and E. Teller, J. Chem. Phys. \textbf{21}, 1087 (1953).

\bibitem {Hastings}W. K. Hastings, Biometrika \textbf{57}, 97 (1970).

\bibitem {SW}R.H. Swendsen and J.-S. Wang, Phys. Rev. Lett. \textbf{58}, 86 (1987).

\bibitem {pottsm}J. -S. Wang, R. H. Swendsen and R. Koteck\'{y}, Phys. Rev.
Lett. \textbf{63}, 1009 (1989).

\bibitem {wolf}U. Wolff, Phys. Rev. Lett. \textbf{62}, 361 (1989).

\bibitem {Thirring}W. Thirring, \textit{Quantum Mechanics of large systems}
(Springer, 1980) Ch. 2.3.

\bibitem {Viana}J. Viana Lopes \textit{et al}, Phys. Rev. E \textbf{74}, 046702
(2006).

\bibitem{BergM}B. Berg and T. Neuhaus, Phys. Lett. B \textbf{267}, 249 (1991); Phys. Rev.
Lett. \textbf{68}, 9 (1992).

\bibitem{WangLandau}F. Wang and D. P. Landau, Phys. Rev. Lett. \textbf{86}, 2050 (2001); Phys.
Rev. E \textbf{64}, 056101 (2001).

\bibitem {Gerling}A. Gerling and R. W. H\"{u}ller, Z. Phys. B \textbf{90}, 207
(1993).

\bibitem{ChallaH}M. S. S. Challa and J. H. Hetherington in \textit{Computer
Simulation Studies in Condensed Matter Physics I}, Eds. D.P. Landau,
K. K. Mon and H.-B. Sch\"{u}ttler (Springer, Heidelberg, 1988).

\bibitem{Hetherington}J. H. Hetherington, J. Low Temp. Phys. \textbf{66}, 145
(1987).

\bibitem {Edwards}R.G. Edwards and A.D. Sokal, Phys. Rev. D \textbf{38}, 2009 (1988).

\bibitem {Niedermayer}F. Niedermayer, Phys. Rev. Lett. 61, 2026 (1988).

\bibitem {Evertz}H.G. Evertz, M. Hasenbusch, M. Marcu, K. Pinn and S. Solomon,
Phys. Lett. B \textbf{254}, 185 (1991).

\bibitem {Hasenbusch}M. Hasenbusch, M. Marcu and K. Pinn, Physica A
\textbf{211}, 255 (1994).

\bibitem {Dress}C. Dress and W. Krauth, J. Phys. A \textbf{28}, L597 (1995).

\bibitem {Liu}J.W. Liu and E. Luijten, Phys. Rev. Lett. \textbf{92}, 035504 (2004).

\bibitem {Wu}F.Y. Wu, Rev. Mod. Phys. \textbf{54}, 235 (1982).

\bibitem {Berg97}B. A. Berg, in \textit{Proceedings of the International Conference
on Multiscale Phenomena and Their Simulations}, Bielefeld, Oct.
1996, Eds. F. Karsch, B. Monien and H. Satz (World Scientific,
Singapore, 1997).

\bibitem {vel-gfdt}L. Velazquez and S. Curilef, \textit{Equilibrium %
fluctuation-dissipation relations: A generalization compatible with %
macrostates with anomalous response functions}, Submited to J. Stat.
Phys. (2009).
\end{thebibliography}
\end{document}